\input harvmac.tex
\input epsf.tex
\input amssym
\input ulem.sty
\input graphicx.tex
%\draftmode

\let\includefigures=\iftrue
\let\useblackboard=\iftrue
\newfam\black

\def\figin{\epsfcheck\figin}\def\figins{\epsfcheck\figins}
\def\epsfcheck{\ifx\epsfbox\UnDeFiNeD
\message{(NO epsf.tex, FIGURES WILL BE IGNORED)}
\gdef\figin##1{\vskip2in}\gdef\figins##1{\hskip.5in}% blank space instead
\else\message{(FIGURES WILL BE INCLUDED)}%
\gdef\figin##1{##1}\gdef\figins##1{##1}\fi}
\def\DefWarn#1{}
\def\figinsert{\goodbreak\midinsert}
\def\ifig#1#2#3{\DefWarn#1\xdef#1{fig.~\the\figno}
\writedef{#1\leftbracket fig.\noexpand~\the\figno} %
\figinsert\figin{\centerline{#3}}\medskip\centerline{\vbox{\baselineskip12pt
\advance\hsize by -1truein\noindent\footnotefont{\bf
Fig.~\the\figno:} #2}}
\bigskip\endinsert\global\advance\figno by1}

% TO INCLUDE FIGURES DO AS BELOW
%\ifig\LABEL{  WRITE CAPTION } {\epsfxsize1.5in\epsfbox{FILENAME.eps}}

%Figure Stuff
\includefigures
\message{If you do not have epsf.tex (to include figures),}
\message{change the option at the top of the tex file.}
\input epsf
\def\figin{\epsfcheck\figin}\def\figins{\epsfcheck\figins}
\def\epsfcheck{\ifx\epsfbox\UnDeFiNeD
\message{(NO epsf.tex, FIGURES WILL BE IGNORED)}
\gdef\figin##1{\vskip2in}\gdef\figins##1{\hskip.5in}% blank space instead
\else\message{(FIGURES WILL BE INCLUDED)}%
\gdef\figin##1{##1}\gdef\figins##1{##1}\fi}
\def\DefWarn#1{}
\def\figinsert{\goodbreak\midinsert}
\def\ifig#1#2#3{\DefWarn#1\xdef#1{fig.~\the\figno}
\writedef{#1\leftbracket fig.\noexpand~\the\figno}%
\figinsert\figin{\centerline{#3}}\medskip\centerline{\vbox{
\baselineskip12pt\advance\hsize by -1truein
\noindent\footnotefont{\bf Fig.~\the\figno:} #2}}
%\bigskip
\endinsert\global\advance\figno by1}
%%%
\else
\def\ifig#1#2#3{\xdef#1{fig.~\the\figno}
\writedef{#1\leftbracket fig.\noexpand~\the\figno}%
%\figinsert\figin{\centerline{#3}}\medskip
%\centerline{\vbox{\baselineskip12pt
%\advance\hsize by -1truein\noindent
%\footnotefont{\bf Fig.~\the\figno:} #2}}
%\bigskip\endinsert
\global\advance\figno by1} \fi

\def\figin{\epsfcheck\figin}\def\figins{\epsfcheck\figins}
\def\epsfcheck{\ifx\epsfbox\UnDeFiNeD
\message{(NO epsf.tex, FIGURES WILL BE IGNORED)}
\gdef\figin##1{\vskip2in}\gdef\figins##1{\hskip.5in}% blank space instead
\else\message{(FIGURES WILL BE INCLUDED)}%
\gdef\figin##1{##1}\gdef\figins##1{##1}\fi}
\def\DefWarn#1{}
\def\figinsert{\goodbreak\midinsert}
\def\ifig#1#2#3{\DefWarn#1\xdef#1{fig.~\the\figno}
\writedef{#1\leftbracket fig.\noexpand~\the\figno} %
\figinsert\figin{\centerline{#3}}\medskip\centerline{\vbox{\baselineskip12pt
\advance\hsize by -1truein\noindent\footnotefont{\bf
Fig.~\the\figno:} #2}}
\bigskip\endinsert\global\advance\figno by1}

\def \pa {\partial}

\def\OO{{\cal OO}}

\catcode`\@=11
\def\slash#1{\mathord{\mathpalette\c@ncel{#1}}}
\overfullrule=0pt
\def\AA{{\cal A}}
\def\BB{{\cal B}}

\def\II{{\cal I}}

\def\OO{{\cal O}}

\def\II{{\cal I}}

\def\underrel#1\over#2{\mathrel{\mathop{\kern\z@#1}\limits_{#2}}}

\catcode`\@=12

%%%%%%%%%%%%%%%%%%%%%%%%%%%%%%%%%%%%%%%%%%%%%%%%%%%%%%%%%%%%%%

\def \cosh{{\rm cosh}}

%%%%%%%%%%%%%%%%%%%%%%%%%%%%%%%%%%%%%%%%%%%%%%%%%%%%%%%%%%%%%%
% new defs:

\def\zbar{{\bar z}}

\def\p{{\partial}}

\def\lra{\leftrightarrow}

\def\hbar{{\bar h}}

\def\DL{{\Delta_{L}}}
\def\DH{{\Delta_{H}}}
\def\DHL{{\Delta_{HL}}}
\def\OL{{\OO_{L}}}
\def\OH{{\OO_{H}}}
\def\CT{{C_{T}}}
\def\ee{{\rm e}}
\def\ebar{{\bar e}}

% References:
%\Afkhami-JeddiNTF
\lref\AfkhamiJeddiNTF{
  N.~Afkhami-Jeddi, T.~Hartman, S.~Kundu and A.~Tajdini,
  ``Einstein gravity 3-point functions from conformal field theory,''
JHEP {\bf 1712}, 049 (2017).
[arXiv:1610.09378 [hep-th]].
%%CITATION = arXiv:1610.09378%%
}
%\AharonyDWX
\lref\AharonyDWX{
  O.~Aharony, L.~F.~Alday, A.~Bissi and E.~Perlmutter,
  ``Loops in AdS from Conformal Field Theory,''
JHEP {\bf 1707}, 036 (2017).
[arXiv:1612.03891 [hep-th]].
%%CITATION = arXiv:1612.03891%%
}
%\AldayGDE
\lref\AldayGDE{
  L.~F.~Alday, A.~Bissi and E.~Perlmutter,
  ``Holographic Reconstruction of AdS Exchanges from Crossing Symmetry,''
JHEP {\bf 1708}, 147 (2017).
[arXiv:1705.02318 [hep-th]].
%%CITATION = arXiv:1705.02318%%
}
%\AldayJFR
\lref\AldayJFR{
  L.~F.~Alday,
  ``Solving CFTs with Weakly Broken Higher Spin Symmetry,''
JHEP {\bf 1710}, 161 (2017).
[arXiv:1612.00696 [hep-th]].
%%CITATION = arXiv:1612.00696%%
}
%\AldayNJK
\lref\AldayNJK{
  L.~F.~Alday,
  ``Large Spin Perturbation Theory for Conformal Field Theories,''
Phys.\ Rev.\ Lett.\  {\bf 119}, no. 11, 111601 (2017).
[arXiv:1611.01500 [hep-th]].
%%CITATION = arXiv:1611.01500%%
}
%\AldayTSA
\lref\AldayTSA{
  L.~F.~Alday, A.~Bissi and T.~Lukowski,
  ``Lessons from crossing symmetry at large N,''
JHEP {\bf 1506}, 074 (2015).
[arXiv:1410.4717 [hep-th]].
%%CITATION = arXiv:1410.4717%%
}

%\CollierEXN
\lref\CollierEXN{
  S.~Collier, Y.~Gobeil, H.~Maxfield and E.~Perlmutter,
  ``Quantum Regge Trajectories and the Virasoro Analytic Bootstrap,''
[arXiv:1811.05710 [hep-th]].
%%CITATION = arXiv:1811.05710%%
}
%\CornalbaAX
\lref\CornalbaAX{
  L.~Cornalba, M.~S.~Costa and J.~Penedones,
  ``Deep Inelastic Scattering in Conformal QCD,''
JHEP {\bf 1003}, 133 (2010).
[arXiv:0911.0043 [hep-th]].
%%CITATION = arXiv:0911.0043%%
}
%\CornalbaXK
\lref\CornalbaXK{
  L.~Cornalba, M.~S.~Costa, J.~Penedones and R.~Schiappa,
  ``Eikonal Approximation in AdS/CFT: From Shock Waves to Four-Point Functions,''
JHEP {\bf 0708}, 019 (2007).
[hep-th/0611122].
%%CITATION = hep-th/0611122%%
}
%\CornalbaXM
\lref\CornalbaXM{
  L.~Cornalba, M.~S.~Costa, J.~Penedones and R.~Schiappa,
  ``Eikonal Approximation in AdS/CFT: Conformal Partial Waves and Finite N Four-Point Functions,''
Nucl.\ Phys.\ B {\bf 767}, 327 (2007).
[hep-th/0611123].
%%CITATION = hep-th/0611123%%
}
%\CornalbaZB
\lref\CornalbaZB{
  L.~Cornalba, M.~S.~Costa and J.~Penedones,
  ``Eikonal approximation in AdS/CFT: Resumming the gravitational loop expansion,''
JHEP {\bf 0709}, 037 (2007).
[arXiv:0707.0120 [hep-th]].
%%CITATION = arXiv:0707.0120%%
}
%\DolanDV
\lref\DolanDV{
  F.~A.~Dolan and H.~Osborn,
  ``Conformal Partial Waves: Further Mathematical Results,''
[arXiv:1108.6194 [hep-th]].
%%CITATION = arXiv:1108.6194%%
}
%\DolanHV
\lref\DolanHV{
  F.~A.~Dolan and H.~Osborn,
  ``Conformal partial waves and the operator product expansion,''
Nucl.\ Phys.\ B {\bf 678}, 491 (2004).
[hep-th/0309180].
%%CITATION = DAMTP-03-91%%
}
%\DolanUT
\lref\DolanUT{
  F.~A.~Dolan and H.~Osborn,
  ``Conformal four point functions and the operator product expansion,''
Nucl.\ Phys.\ B {\bf 599}, 459 (2001).
[hep-th/0011040].
%%CITATION = hep-th/0011040%%
}
%\ElShowkHT
\lref\ElShowkHT{
  S.~El-Showk, M.~F.~Paulos, D.~Poland, S.~Rychkov, D.~Simmons-Duffin and A.~Vichi,
  ``Solving the 3D Ising Model with the Conformal Bootstrap,''
Phys.\ Rev.\ D {\bf 86}, 025022 (2012).
[arXiv:1203.6064 [hep-th]].
%%CITATION = LPTENS-12-07%%
}
%\FerraraYT
\lref\FerraraYT{
  S.~Ferrara, A.~F.~Grillo and R.~Gatto,
  ``Tensor representations of conformal algebra and conformally covariant operator product expansion,''
Annals Phys.\  {\bf 76}, 161 (1973)..
}
%\FitzpatrickDLT
\lref\FitzpatrickDLT{
  A.~L.~Fitzpatrick and J.~Kaplan,
  ``Conformal Blocks Beyond the Semi-Classical Limit,''
JHEP {\bf 1605}, 075 (2016).
[arXiv:1512.03052 [hep-th]].
%%CITATION = arXiv:1512.03052%%
}
%\FitzpatrickDM
\lref\FitzpatrickDM{
  A.~L.~Fitzpatrick and J.~Kaplan,
  ``Unitarity and the Holographic S-Matrix,''
JHEP {\bf 1210}, 032 (2012).
[arXiv:1112.4845 [hep-th]].
%%CITATION = SLAC-PUB-14979%%
}
%\FitzpatrickZHA
\lref\FitzpatrickZHA{
  A.~L.~Fitzpatrick, J.~Kaplan and M.~T.~Walters,
  ``Virasoro Conformal Blocks and Thermality from Classical Background Fields,''
JHEP {\bf 1511}, 200 (2015).
[arXiv:1501.05315 [hep-th]].
%%CITATION = arXiv:1501.05315%%
}

%\FitzpatrickIVE
\lref\FitzpatrickIVE{
  A.~L.~Fitzpatrick, J.~Kaplan, D.~Li and J.~Wang,
  ``On information loss in AdS$_{3}$/CFT$_{2}$,''
JHEP {\bf 1605}, 109 (2016).
[arXiv:1603.08925 [hep-th]].
%%CITATION = arXiv:1603.08925%%
}
%\FitzpatrickMJQ
\lref\FitzpatrickMJQ{
  A.~L.~Fitzpatrick and J.~Kaplan,
  ``On the Late-Time Behavior of Virasoro Blocks and a Classification of Semiclassical Saddles,''
JHEP {\bf 1704}, 072 (2017).
[arXiv:1609.07153 [hep-th]].
%%CITATION = arXiv:1609.07153%%
}
%\FitzpatrickVUA
\lref\FitzpatrickVUA{
  A.~L.~Fitzpatrick, J.~Kaplan and M.~T.~Walters,
  ``Universality of Long-Distance AdS Physics from the CFT Bootstrap,''
JHEP {\bf 1408}, 145 (2014).
[arXiv:1403.6829 [hep-th]].
%%CITATION = arXiv:1403.6829%%
}
%\FitzpatrickYX
\lref\FitzpatrickYX{
  A.~L.~Fitzpatrick, J.~Kaplan, D.~Poland and D.~Simmons-Duffin,
  ``The Analytic Bootstrap and AdS Superhorizon Locality,''
JHEP {\bf 1312}, 004 (2013).
[arXiv:1212.3616 [hep-th]].
%%CITATION = arXiv:1212.3616%%
}
%\FitzpatrickZHA
\lref\FitzpatrickZHA{
  A.~L.~Fitzpatrick, J.~Kaplan and M.~T.~Walters,
  ``Virasoro Conformal Blocks and Thermality from Classical Background Fields,''
JHEP {\bf 1511}, 200 (2015).
[arXiv:1501.05315 [hep-th]].
%%CITATION = arXiv:1501.05315%%
}
%\GubserBC
\lref\GubserBC{
  S.~S.~Gubser, I.~R.~Klebanov and A.~M.~Polyakov,
  ``Gauge theory correlators from noncritical string theory,''
Phys.\ Lett.\ B {\bf 428}, 105 (1998).
[hep-th/9802109].
%%CITATION = hep-th/9802109%%
}
%\HartmanLFA
\lref\HartmanLFA{
  T.~Hartman, S.~Jain and S.~Kundu,
  ``Causality Constraints in Conformal Field Theory,''
JHEP {\bf 1605}, 099 (2016).
[arXiv:1509.00014 [hep-th]].
%%CITATION = arXiv:1509.00014%%
}
%\HeemskerkTY
\lref\HeemskerkTY{
  I.~Heemskerk and J.~Sully,
  ``More Holography from Conformal Field Theory,''
JHEP {\bf 1009}, 099 (2010).
[arXiv:1006.0976 [hep-th]].
%%CITATION = arXiv:1006.0976%%
}
%\HeemskerkPN
\lref\HeemskerkPN{
  I.~Heemskerk, J.~Penedones, J.~Polchinski and J.~Sully,
  ``Holography from Conformal Field Theory,''
JHEP {\bf 0910}, 079 (2009).
[arXiv:0907.0151 [hep-th]].
%%CITATION = arXiv:0907.0151%%
}
%\HofmanAWC
\lref\HofmanAWC{
  D.~M.~Hofman, D.~Li, D.~Meltzer, D.~Poland and F.~Rejon-Barrera,
  ``A Proof of the Conformal Collider Bounds,''
JHEP {\bf 1606}, 111 (2016).
[arXiv:1603.03771 [hep-th]].
%%CITATION = arXiv:1603.03771%%
}
%\KavirajCXA
\lref\KavirajCXA{
  A.~Kaviraj, K.~Sen and A.~Sinha,
  ``Analytic bootstrap at large spin,''
JHEP {\bf 1511}, 083 (2015).
[arXiv:1502.01437 [hep-th]].
%%CITATION = arXiv:1502.01437%%
}
%\KavirajXSA
\lref\KavirajXSA{
  A.~Kaviraj, K.~Sen and A.~Sinha,
  ``Universal anomalous dimensions at large spin and large twist,''
JHEP {\bf 1507}, 026 (2015).
[arXiv:1504.00772 [hep-th]].
%%CITATION = arXiv:1504.00772%%
}
%\KomargodskiEK
\lref\KomargodskiEK{
  Z.~Komargodski and A.~Zhiboedov,
  ``Convexity and Liberation at Large Spin,''
JHEP {\bf 1311}, 140 (2013).
[arXiv:1212.4103 [hep-th]].
%%CITATION = arXiv:1212.4103%%
}
%\KomargodskiGCI
\lref\KomargodskiGCI{
  Z.~Komargodski, M.~Kulaxizi, A.~Parnachev and A.~Zhiboedov,
  ``Conformal Field Theories and Deep Inelastic Scattering,''
Phys.\ Rev.\ D {\bf 95}, no. 6, 065011 (2017).
[arXiv:1601.05453 [hep-th]].
%%CITATION = arXiv:1601.05453%%
}
%\KulaxiziDXO
\lref\KulaxiziDXO{
  M.~Kulaxizi, G.~S.~Ng and A.~Parnachev,
  ``Black Holes, Heavy States, Phase Shift and Anomalous Dimensions,''
[arXiv:1812.03120 [hep-th]].
%%CITATION = arXiv:1812.03120%%
}
%\KulaxiziIXA
\lref\KulaxiziIXA{
  M.~Kulaxizi, A.~Parnachev and A.~Zhiboedov,
  ``Bulk Phase Shift, CFT Regge Limit and Einstein Gravity,''
JHEP {\bf 1806}, 121 (2018).
[arXiv:1705.02934 [hep-th]].
%%CITATION = arXiv:1705.02934%%
}
%\LiLMH
\lref\LiLMH{
  D.~Li, D.~Meltzer and D.~Poland,
  ``Conformal Bootstrap in the Regge Limit,''
JHEP {\bf 1712}, 013 (2017).
[arXiv:1705.03453 [hep-th]].
%%CITATION = arXiv:1705.03453%%
}
%\LiuJHS
\lref\LiuJHS{
  J.~Liu, E.~Perlmutter, V.~Rosenhaus and D.~Simmons-Duffin,
  ``$d$-dimensional SYK, AdS Loops, and $6j$ Symbols,''
[arXiv:1808.00612 [hep-th]].
%%CITATION = arXiv:1808.00612%%
}

%\MaldacenaRE
\lref\MaldacenaRE{
  J.~M.~Maldacena,
  ``The Large N limit of superconformal field theories and supergravity,''
Int.\ J.\ Theor.\ Phys.\  {\bf 38}, 1113 (1999), [Adv.\ Theor.\ Math.\ Phys.\  {\bf 2}, 231 (1998)].
[hep-th/9711200].
%%CITATION = HUTP-97-A097%%
}

%\PolandEPD
\lref\PolandEPD{
  D.~Poland, S.~Rychkov and A.~Vichi,
  ``The Conformal Bootstrap: Theory, Numerical Techniques, and Applications,''
[arXiv:1805.04405 [hep-th]].
%%CITATION = arXiv:1805.04405%%
}
%\PolyakovGS
\lref\PolyakovGS{
  A.~M.~Polyakov,
  ``Nonhamiltonian approach to conformal quantum field theory,''
Zh.\ Eksp.\ Teor.\ Fiz.\  {\bf 66}, 23 (1974), [Sov.\ Phys.\ JETP {\bf 39}, 9 (1974)]..
}
%\RattazziPE
\lref\RattazziPE{
  R.~Rattazzi, V.~S.~Rychkov, E.~Tonni and A.~Vichi,
  ``Bounding scalar operator dimensions in 4D CFT,''
JHEP {\bf 0812}, 031 (2008).
[arXiv:0807.0004 [hep-th]].
%%CITATION = arXiv:0807.0004%%
}
%\RychkovIQZ
\lref\RychkovIQZ{
  S.~Rychkov,
  ``EPFL Lectures on Conformal Field Theory in D>= 3 Dimensions,''
[arXiv:1601.05000 [hep-th]].
%%CITATION = CERN-TH-2016-012%%
}
%\Simmons-DuffinGJK
\lref\SimmonsDuffinGJK{
  D.~Simmons-Duffin,
  ``The Conformal Bootstrap,''
[arXiv:1602.07982 [hep-th]].
%%CITATION = arXiv:1602.07982%%
}

%\WittenQJ
\lref\WittenQJ{
  E.~Witten,
  ``Anti-de Sitter space and holography,''
Adv.\ Theor.\ Math.\ Phys.\  {\bf 2}, 253 (1998).
[hep-th/9802150].
%%CITATION = hep-th/9802150%%
}

%\ElShowkAG
\lref\ElShowkAG{
  S.~El-Showk and K.~Papadodimas,
  ``Emergent Spacetime and Holographic CFTs,''
JHEP {\bf 1210}, 106 (2012).
[arXiv:1101.4163 [hep-th]].
%%CITATION = arXiv:1101.4163%%
}

%\FitzpatrickIA
\lref\FitzpatrickIA{
  A.~L.~Fitzpatrick, J.~Kaplan, J.~Penedones, S.~Raju and B.~C.~van Rees,
  ``A Natural Language for AdS/CFT Correlators,''
JHEP {\bf 1111}, 095 (2011).
[arXiv:1107.1499 [hep-th]].
%%CITATION = SLAC-PUB-14506%%
}

%\FitzpatrickHU
\lref\FitzpatrickHU{
  A.~L.~Fitzpatrick and J.~Kaplan,
  ``Analyticity and the Holographic S-Matrix,''
JHEP {\bf 1210}, 127 (2012).
[arXiv:1111.6972 [hep-th]].
%%CITATION = SLAC-PUB-14841%%
}

%\FitzpatrickDM
\lref\FitzpatrickDM{
  A.~L.~Fitzpatrick and J.~Kaplan,
  ``Unitarity and the Holographic S-Matrix,''
JHEP {\bf 1210}, 032 (2012).
[arXiv:1112.4845 [hep-th]].
%%CITATION = SLAC-PUB-14979%%
}

%\FitzpatrickCG
\lref\FitzpatrickCG{
  A.~L.~Fitzpatrick and J.~Kaplan,
  ``AdS Field Theory from Conformal Field Theory,''
JHEP {\bf 1302}, 054 (2013).
[arXiv:1208.0337 [hep-th]].
%%CITATION = arXiv:1208.0337%%
}

%\FitzpatrickVUA
\lref\FitzpatrickVUA{
  A.~L.~Fitzpatrick, J.~Kaplan and M.~T.~Walters,
  ``Universality of Long-Distance AdS Physics from the CFT Bootstrap,''
JHEP {\bf 1408}, 145 (2014).
[arXiv:1403.6829 [hep-th]].
%%CITATION = arXiv:1403.6829%%
}

%\AldayTSA
\lref\AldayTSA{
  L.~F.~Alday, A.~Bissi and T.~Lukowski,
  ``Lessons from crossing symmetry at large N,''
JHEP {\bf 1506}, 074 (2015).
[arXiv:1410.4717 [hep-th]].
%%CITATION = arXiv:1410.4717%%
}

%\HijanoZSA
\lref\HijanoZSA{
  E.~Hijano, P.~Kraus, E.~Perlmutter and R.~Snively,
  ``Witten Diagrams Revisited: The AdS Geometry of Conformal Blocks,''
JHEP {\bf 1601}, 146 (2016).
[arXiv:1508.00501 [hep-th]].
%%CITATION = arXiv:1508.00501%%
}

%\MaldacenaIUA
\lref\MaldacenaIUA{
  J.~Maldacena, D.~Simmons-Duffin and A.~Zhiboedov,
  ``Looking for a bulk point,''
JHEP {\bf 1701}, 013 (2017).
[arXiv:1509.03612 [hep-th]].
%%CITATION = arXiv:1509.03612%%
}

%\HenrikssonEEJ
\lref\HenrikssonEEJ{
  J.~Henriksson and T.~Lukowski,
  ``Perturbative Four-Point Functions from the Analytic Conformal Bootstrap,''
JHEP {\bf 1802}, 123 (2018).
[arXiv:1710.06242 [hep-th]].
%%CITATION = arXiv:1710.06242%%
}

%\AldayVKK
\lref\AldayVKK{
  L.~F.~Alday and S.~Caron-Huot,
  ``Gravitational S-matrix from CFT dispersion relations,''
JHEP {\bf 1812}, 017 (2018).
[arXiv:1711.02031 [hep-th]].
%%CITATION = arXiv:1711.02031%%
}

%\Afkhami-JeddiRMX
\lref\AfkhamiJeddiRMX{
  N.~Afkhami-Jeddi, T.~Hartman, S.~Kundu and A.~Tajdini,
  ``Shockwaves from the Operator Product Expansion,''
[arXiv:1709.03597 [hep-th]].
%%CITATION = arXiv:1709.03597%%
}

%\CostaTWZ
\lref\CostaTWZ{
  M.~S.~Costa, T.~Hansen and J.~Penedones,
  ``Bounds for OPE coefficients on the Regge trajectory,''
JHEP {\bf 1710}, 197 (2017).
[arXiv:1707.07689 [hep-th]].
%%CITATION = arXiv:1707.07689%%
}

%\VosPQA
\lref\VosPQA{
  G.~Vos,
  ``Generalized Additivity in Unitary Conformal Field Theories,''
Nucl.\ Phys.\ B {\bf 899}, 91 (2015).
[arXiv:1411.7941 [hep-th]].
%%CITATION = arXiv:1411.7941%%
}

%\AldayEYA
\lref\AldayEYA{
  L.~F.~Alday, A.~Bissi and T.~Lukowski,
  ``Large spin systematics in CFT,''
JHEP {\bf 1511}, 101 (2015).
[arXiv:1502.07707 [hep-th]].
%%CITATION = arXiv:1502.07707%%
}

%\DeyZBG
\lref\DeyZBG{
  P.~Dey, A.~Kaviraj and K.~Sen,
  ``More on analytic bootstrap for O(N) models,''
JHEP {\bf 1606}, 136 (2016).
[arXiv:1602.04928 [hep-th]].
%%CITATION = arXiv:1602.04928%%
}

%\AharonyNPF
\lref\AharonyNPF{
  O.~Aharony, L.~F.~Alday, A.~Bissi and R.~Yacoby,
  ``The Analytic Bootstrap for Large $N$ Chern-Simons Vector Models,''
JHEP {\bf 1808}, 166 (2018).
[arXiv:1805.04377 [hep-th]].
%%CITATION = arXiv:1805.04377%%
}

%\SleightRYU
\lref\SleightRYU{
  C.~Sleight and M.~Taronna,
  ``Anomalous Dimensions from Crossing Kernels,''
JHEP {\bf 1811}, 089 (2018).
[arXiv:1807.05941 [hep-th]].
%%CITATION = arXiv:1807.05941%%
}

%\CardonaDOV
\lref\CardonaDOV{
  C.~Cardona and K.~Sen,
  ``Anomalous dimensions at finite conformal spin from OPE inversion,''
JHEP {\bf 1811}, 052 (2018).
[arXiv:1806.10919 [hep-th]].
%%CITATION = arXiv:1806.10919%%
}

%\SleightEPI
\lref\SleightEPI{
  C.~Sleight and M.~Taronna,
  ``Spinning Mellin Bootstrap: Conformal Partial Waves, Crossing Kernels and Applications,''
Fortsch.\ Phys.\  {\bf 66}, no. 8-9, 1800038 (2018).
[arXiv:1804.09334 [hep-th]].
%%CITATION = arXiv:1804.09334%%
}

%\FitzpatrickZQZ
\lref\FitzpatrickZQZ{
  A.~L.~Fitzpatrick and K.~W.~Huang,
  ``Universal Lowest-Twist in CFTs from Holography,''
[arXiv:1903.05306 [hep-th]].
%%CITATION = arXiv:1903.05306%%
}

%\BrowerEA
\lref\BrowerEA{
  R.~C.~Brower, J.~Polchinski, M.~J.~Strassler and C.~I.~Tan,
  ``The Pomeron and gauge/string duality,''
JHEP {\bf 0712}, 005 (2007).
[hep-th/0603115].
%%CITATION = hep-th/0603115%%
}
%\ShenkerCWA
\lref\ShenkerCWA{
  S.~H.~Shenker and D.~Stanford,
  ``Stringy effects in scrambling,''
JHEP {\bf 1505}, 132 (2015).
[arXiv:1412.6087 [hep-th]].
%%CITATION = arXiv:1412.6087%%
}
%\GallianiCAI
\lref\GallianiCAI{
  A.~Galliani, S.~Giusto, E.~Moscato and R.~Russo,
  ``Correlators at large c without information loss,''
JHEP {\bf 1609}, 065 (2016).
[arXiv:1606.01119 [hep-th]].
%%CITATION = QMUL-PH-16-12%%
}
%\FitzpatrickZQZ
\lref\FitzpatrickZQZ{
  A.~L.~Fitzpatrick and K.~W.~Huang,
  ``Universal Lowest-Twist in CFTs from Holography,''
[arXiv:1903.05306 [hep-th]].
%%CITATION = arXiv:1903.05306%%
}
%\FitzpatrickMJQ
\lref\FitzpatrickMJQ{
  A.~L.~Fitzpatrick and J.~Kaplan,
  ``On the Late-Time Behavior of Virasoro Blocks and a Classification of Semiclassical Saddles,''
JHEP {\bf 1704}, 072 (2017).
[arXiv:1609.07153 [hep-th]].
%%CITATION = arXiv:1609.07153%%
}

%\CostaSCB
\lref\CostaSCB{
  M.~S.~Costa, J.~Penedones, D.~Poland and S.~Rychkov
  ``Spinning Conformal Blocks,''
JHEP {\bf 1111}, 154 (2015).
[arXiv:1109.6321 [hep-th]].
%%CITATION = arXiv:1109.6321%%
}

%\SimmonsDuffinUY
\lref\SimmonsDuffinUY{
  D.~Simmons-Duffin,
  ``Projectors, Shadows, and Conformal Blocks,''
JHEP {\bf 1404}, 146 (2014).
[arXiv:1204.3894 [hep-th]].
%%CITATION = arXiv:1204.3894%%
}

%\CornalbaFS
\lref\CornalbaFS{
  L.~Cornalba,
  ''Eikonal methods in AdS/CFT: Regge theory and multi-reggeon exchange,''
[arXiv:0710.5480 [hep-th]].
%%CITATION = arXiv:0710.5480%%
}

%\CostaCB
\lref\CostaCB{
  M.~S.~Costa, V.~Goncalves and J.~Penedones,
  ``Conformal Regge theory,''
JHEP {\bf 1212}, 091 (2012).
[arXiv:1209.4355 [hep-th]].
%%CITATION = arXiv:1209.4355%%
}

%\BanksBJ
\lref\BanksBJ{
  T.~Banks and G.~Festuccia,
  ``The Regge Limit for Green Functions in Conformal Field Theory,''
JHEP {\bf 1006}, 105 (2010).
[arXiv:0910.2746 [hep-th]].
%%CITATION = SCIPP-2009-15%%
}

%\GiordanoUA
\lref\GiordanoUA{
  M.~Giordano, R.~Peschanski and S.~Seki,
  ``Eikonal Approach to N=4 SYM Regge Amplitudes in the AdS/CFT Correspondence,''
Acta Phys.\ Polon.\ B {\bf 43}, 1289 (2012).
[arXiv:1110.3680 [hep-th]].
%%CITATION = arXiv:1110.3680%%
}

%\CostaZRA
\lref\CostaZRA{
  M.~S.~Costa, J.~Drummond, V.~Goncalves and J.~Penedones,
  ``The role of leading twist operators in the Regge and Lorentzian OPE limits,''
JHEP {\bf 1404}, 094 (2014).
[arXiv:1311.4886 [hep-th]].
%%CITATION = arXiv:1311.4886%%
}

%\Caron-HuotVEP
\lref\CaronHuotVEP{
  S.~Caron-Huot,
  ``Analyticity in Spin in Conformal Theories,''
JHEP {\bf 1709}, 078 (2017).
[arXiv:1703.00278 [hep-th]].
%%CITATION = arXiv:1703.00278%%
}

%\Afkhami-JeddiRMX
\lref\AfkhamiJeddiRMX{
  N.~Afkhami-Jeddi, T.~Hartman, S.~Kundu and A.~Tajdini,
  ``Shockwaves from the Operator Product Expansion,''
[arXiv:1709.03597 [hep-th]].
%%CITATION = arXiv:1709.03597%%
}

%\Simmons-DuffinNUB
\lref\SimmonsDuffinNUB{
  D.~Simmons-Duffin, D.~Stanford and E.~Witten,
  ``A spacetime derivation of the Lorentzian OPE inversion formula,''
JHEP {\bf 1807}, 085 (2018).
[arXiv:1711.03816 [hep-th]].
%%CITATION = arXiv:1711.03816%%
}

%\JafferisZNA
\lref\JafferisZNA{
  D.~Jafferis, B.~Mukhametzhanov and A.~Zhiboedov,
  ``Conformal Bootstrap At Large Charge,''
JHEP {\bf 1805}, 043 (2018).
[arXiv:1710.11161 [hep-th]].
%%CITATION = arXiv:1710.11161%%
}

%\MeltzerRTF
\lref\MeltzerRTF{
  D.~Meltzer and E.~Perlmutter,
  ``Beyond $a = c$: gravitational couplings to matter and the stress tensor OPE,''
JHEP {\bf 1807}, 157 (2018).
[arXiv:1712.04861 [hep-th]].
%%CITATION = arXiv:1712.04861%%
}

%\LemosVNX
\lref\LemosVNX{
  M.~Lemos, P.~Liendo, M.~Meineri and S.~Sarkar,
  ``Universality at large transverse spin in defect CFT,''
JHEP {\bf 1809}, 091 (2018).
[arXiv:1712.08185 [hep-th]].
%%CITATION = arXiv:1712.08185%%
}

%\IliesiuFAO
\lref\IliesiuFAO{
  L.~Iliesiu, M.~Kologlu, R.~Mahajan, E.~Perlmutter and D.~Simmons-Duffin,
  ``The Conformal Bootstrap at Finite Temperature,''
JHEP {\bf 1810}, 070 (2018).
[arXiv:1802.10266 [hep-th]].
%%CITATION = arXiv:1802.10266%%
}

%\KravchukHTV
\lref\KravchukHTV{
  P.~Kravchuk and D.~Simmons-Duffin,
  ``Light-ray operators in conformal field theory,''
JHEP {\bf 1811}, 102 (2018).
[arXiv:1805.00098 [hep-th]].
%%CITATION = arXiv:1805.00098%%
}

%\Afkhami-JeddiOWN
\lref\AfkhamiJeddiOWN{
  N.~Afkhami-Jeddi, S.~Kundu and A.~Tajdini,
  ``A Conformal Collider for Holographic CFTs,''
JHEP {\bf 1810}, 156 (2018).
[arXiv:1805.07393 [hep-th]].
%%CITATION = arXiv:1805.07393%%
}

%\MeltzerTNM
\lref\MeltzerTNM{
  D.~Meltzer,
  ``Higher Spin ANEC and the Space of CFTs,''
[arXiv:1811.01913 [hep-th]].
%%CITATION = arXiv:1811.01913%%
}

%\Afkhami-JeddiAPJ
\lref\AfkhamiJeddiAPJ{
  N.~Afkhami-Jeddi, S.~Kundu and A.~Tajdini,
  ``A Bound on Massive Higher Spin Particles,''
[arXiv:1811.01952 [hep-th]].
%%CITATION = arXiv:1811.01952%%
}

%\CardonaYMB
\lref\CardonaYMB{
  C.~Cardona, Y.~t.~Huang and T.~H.~Tsai,
  ``On the linearity of Regge trajectory at large transfer energy,''
[arXiv:1611.05797 [hep-th]].
%%CITATION = arXiv:1611.05797%%
}

%\AldayHTQ
\lref\AldayHTQ{
  L.~F.~Alday and A.~Bissi,
  ``Unitarity and positivity constraints for CFT at large central charge,''
JHEP {\bf 1707}, 044 (2017).
[arXiv:1606.09593 [hep-th]].
%%CITATION = arXiv:1606.09593%%
}

%\BartelsSC
\lref\BartelsSC{
  J.~Bartels, J.~Kotanski, A.-M.~Mischler and V.~Schomerus,
  ``Regge limit of R-current correlators in AdS Supergravity,''
Nucl.\ Phys.\ B {\bf 830}, 153 (2010).
[arXiv:0908.2301 [hep-th]].
%%CITATION = DESY-09-118%%
}

%\BartelsZY
\lref\BartelsZY{
  J.~Bartels, A.-M.~Mischler and M.~Salvadore,
  ``Four point function of R-currents in N=4 SYM in the Regge limit at weak coupling,''
Phys.\ Rev.\ D {\bf 78}, 016004 (2008).
[arXiv:0803.1423 [hep-ph]].
%%CITATION = arXiv:0803.1423%%
}

%\BeccariaSHQ
\lref\BeccariaSHQ{
  M.~Beccaria, A.~Fachechi and G.~Macorini,
  ``Virasoro vacuum block at next-to-leading order in the heavy-light limit,''
JHEP {\bf 1602}, 072 (2016).
[arXiv:1511.05452 [hep-th]].
%%CITATION = arXiv:1511.05452%%
}

%\ChenCMS
\lref\ChenCMS{
  H.~Chen, A.~L.~Fitzpatrick, J.~Kaplan, D.~Li and J.~Wang,
  ``Degenerate Operators and the $1/c$ Expansion: Lorentzian Resummations, High Order Computations, and Super-Virasoro Blocks,''
JHEP {\bf 1703}, 167 (2017).
[arXiv:1606.02659 [hep-th]].
%%CITATION = arXiv:1606.02659%%
}

%\RastelliUDC
\lref\RastelliUDC{
  L.~Rastelli and X.~Zhou,
  ``How to Succeed at Holographic Correlators Without Really Trying,''
JHEP {\bf 1804}, 014 (2018).
[arXiv:1710.05923 [hep-th]].
%%CITATION = YITP-SB-2017-44%%
}

%\FitzpatrickZM
\lref\FitzpatrickZM{
  A.~L.~Fitzpatrick, E.~Katz, D.~Poland and D.~Simmons-Duffin,
  ``Effective Conformal Theory and the Flat-Space Limit of AdS,''
JHEP {\bf 1107}, 023 (2011).
[arXiv:1007.2412 [hep-th]].
%%CITATION = arXiv:1007.2412%%
}

%\PenedonesUE
\lref\PenedonesUE{
  J.~Penedones,
  ``Writing CFT correlation functions as AdS scattering amplitudes,''
JHEP {\bf 1103}, 025 (2011).
[arXiv:1011.1485 [hep-th]].
%%CITATION = arXiv:1011.1485%%
}

%\GoncalvesRFA
\lref\GoncalvesRFA{
  V.~Gonçalves, J.~Penedones and E.~Trevisani,
  ``Factorization of Mellin amplitudes,''
JHEP {\bf 1510}, 040 (2015).
[arXiv:1410.4185 [hep-th]].
%%CITATION = arXiv:1410.4185%%
}

%\AldayHTQ
\lref\AldayHTQ{
  L.~F.~Alday and A.~Bissi,
  ``Unitarity and positivity constraints for CFT at large central charge,''
JHEP {\bf 1707}, 044 (2017).
[arXiv:1606.09593 [hep-th]].
%%CITATION = arXiv:1606.09593%%
}

%\AprileBGS
\lref\AprileBGS{
  F.~Aprile, J.~M.~Drummond, P.~Heslop and H.~Paul,
  ``Quantum Gravity from Conformal Field Theory,''
JHEP {\bf 1801}, 035 (2018).
[arXiv:1706.02822 [hep-th]].
%%CITATION = arXiv:1706.02822%%
}

%\RastelliNZE
\lref\RastelliNZE{
  L.~Rastelli and X.~Zhou,
  ``Mellin amplitudes for $AdS_5\times S^5$,''
Phys.\ Rev.\ Lett.\  {\bf 118}, no. 9, 091602 (2017).
[arXiv:1608.06624 [hep-th]].
%%CITATION = arXiv:1608.06624%%
}

%\AldayXUA
\lref\AldayXUA{
  L.~F.~Alday and A.~Bissi,
  ``Loop Corrections to Supergravity on $AdS_5 \times S^5$,''
Phys.\ Rev.\ Lett.\  {\bf 119}, no. 17, 171601 (2017).
[arXiv:1706.02388 [hep-th]].
%%CITATION = arXiv:1706.02388%%
}

%\AldayNJK
\lref\AldayNJK{
  L.~F.~Alday,
  ``Large Spin Perturbation Theory for Conformal Field Theories,''
Phys.\ Rev.\ Lett.\  {\bf 119}, no. 11, 111601 (2017).
[arXiv:1611.01500 [hep-th]].
%%CITATION = arXiv:1611.01500%%
}

%\MaldacenaWAA
\lref\MaldacenaWAA{
  J.~Maldacena, S.~H.~Shenker and D.~Stanford,
  ``A bound on chaos,''
JHEP {\bf 1608}, 106 (2016).
[arXiv:1503.01409 [hep-th]].
%%CITATION = arXiv:1503.01409%%
}

%\BrowerEA
\lref\BrowerEA{
  R.~C.~Brower, J.~Polchinski, M.~J.~Strassler and C.~I.~Tan,
  ``The Pomeron and gauge/string duality,''
JHEP {\bf 0712}, 005 (2007).
[hep-th/0603115].
%%CITATION = hep-th/0603115%%
}

%\CardonaQRT
\lref\CardonaQRT{
  C.~Cardona, S.~Guha, S.~K.~Kanumilli and K.~Sen,
  ``Resummation at finite conformal spin,''
JHEP {\bf 1901}, 077 (2019).
[arXiv:1811.00213 [hep-th]].
%%CITATION = arXiv:1811.00213%%
}

%\PaulosSMB
\lref\PaulosSMB{
  M.~F.~Paulos, J.~Penedones, J.~Toledo, B.~C.~van Rees and P.~Vieira,
  ``The S-matrix Bootstrap I: QFT in AdS,''
[arXiv:1607.06109 [hep-th]].
%%CITATION = arXiv:1811.00213%%
}

\Title{
\vbox{\baselineskip8pt
% \hbox{SPIN-07/41} \hbox{
%ITP-UU-07/55}
}}
{\vbox{
\centerline{Black Holes and Conformal Regge Bootstrap }
%\vskip.1in
%\centerline{operators in the Regge limit}
}}

\vskip.1in
 \centerline{
Robin Karlsson, Manuela Kulaxizi, Andrei Parnachev and Petar Tadi\' c } \vskip.1in
\centerline{\it 
School of Mathematics, Trinity College Dublin, Dublin 2, Ireland}

\vskip.7in \centerline{\bf Abstract}{
\vskip.2in 
Highly energetic particles  traveling in the background of 
an asymptotically AdS black hole experience a Shapiro time delay and an angle deflection.
These quantities are related to the Regge limit of a  heavy-heavy-light-light four-point function of scalar operators in the dual CFT.
The Schwarzschild radius of the black hole  in AdS units is proportional to the ratio of the conformal dimension
of the heavy operator and the central charge. 
This ratio serves as a useful expansion parameter; its
power counts the number of  stress tensors in the multi-stress tensor operators
which contribute to the four-point function.
In the cross-channel  the four-point function is determined by  the OPE coefficients
and anomalous dimensions of the heavy-light double-trace operators.
We explain how this data can be obtained and explicitly compute the first and second order terms in the expansion  of the anomalous dimensions.
We observe perfect agreement with known results in the lightcone limit, which were obtained by computing perturbative
corrections to the energy eigenstates in AdS spacetimes.
}

\Date{March 2019}

\listtoc\writetoc
\vskip 1.57in \noindent

\eject
%\newpage

%\draftmode

\newsec{Introduction and Summary}
\noindent

\subsec{Introduction}

\noindent The AdS/CFT correspondence provides a non-perturbative definition of quantum gravity in negatively curved spacetimes \refs{\MaldacenaRE\WittenQJ-\GubserBC}. 
%Since conformal field theories are well-defined non-perturbatively, 
The correspondence in principle provides an opportunity to study generic properties of quantum gravity, possibly probing regimes unattainable by low-energy effective theories. 
%Moreover, one does not necessarily need to make any assumption about the gravity theory being described by, e.g., Einstein gravity. 
Recent years have seen a development in conformal bootstrap techniques following \refs{\FerraraYT\PolyakovGS\RattazziPE-\ElShowkHT}, leading to many results for 
%general CFTs \refs{\AldayJFR\AldayNJK\HofmanAWC\CaronHuotVEP\SimmonsDuffinNUB\LiuJHS,-\HartmanLFA} as well as for 
CFTs with holographic duals  (see e.g.
\refs{\HeemskerkPN\HeemskerkTY\FitzpatrickZM\PenedonesUE\ElShowkAG\FitzpatrickIA\FitzpatrickHU\FitzpatrickDM\FitzpatrickCG\FitzpatrickVUA\GoncalvesRFA\AldayTSA\HijanoZSA\MaldacenaIUA\AldayHTQ\RastelliNZE\AharonyDWX\AldayNJK\AldayGDE\AldayXUA\AprileBGS\RastelliUDC-\AldayVKK}).
CFT methods have therefore become a powerful tool in the study of quantum gravity.

Crossing symmetry in CFTs imposes highly non-trivial constraints on the theory. The idea of conformal bootstrap is to use these constraints to put restrictions on the CFT data or, if possible, even solve the theory. 
%In practice this is in most cases beyond our reach. 
One way to make use of the crossing symmetry  is to isolate a small number of contributing operators in one channel, e.g.\ by going to a certain kinematical regime. This typically has to be reproduced by the exchange of an infinite number of operators in another channel. One such example is the lightcone limit where the separation between two operators
in a four-point function is close to being null.
One can then infer  \refs{\FitzpatrickYX,\KomargodskiEK} the existence of double-twist operators at large spin in any CFT 
in dimensions $d>2$.
The Regge limit provides another opportunity to isolate the contribution of a class of operators, those of highest spin. 
%in one channel, making CFTs with a low number of higher spin operators tractable to study in this limit. 

Holographic CFTs satisfy the following defining properties: (1) large central charge $C_T\sim N^2$ and large-$N$ factorization of correlation functions and (2) a parametrically large gap in the spectrum of single trace operators above spin-$2$. As argued in
\HeemskerkPN, they are dual to theories of quantum gravity in asymptotically AdS spacetimes with local physics below the AdS scale. 
In holographic CFTs the Regge limit of a four-point function, extensively studied in \refs{\CornalbaXK\CornalbaXM\CornalbaZB\CornalbaFS\CornalbaAX-\CostaCB}\foot{See also 
\refs{\BrowerEA\BartelsZY\BartelsSC\BanksBJ\GiordanoUA\CostaZRA\AldayHTQ\MaldacenaWAA\AfkhamiJeddiNTF\CardonaYMB\CaronHuotVEP\KulaxiziIXA\LiLMH\CostaTWZ\AfkhamiJeddiRMX\SimmonsDuffinNUB\JafferisZNA\MeltzerRTF\LemosVNX\IliesiuFAO\KravchukHTV\AfkhamiJeddiOWN\MeltzerTNM-\AfkhamiJeddiAPJ
} for other recent applications of Regge limit in CFTs.}, is dominated by operators of spin two -- the stress tensor and
the double-trace operators (this is a consequence of the gap in the spectrum).
In gravity, it reproduces a Witten diagram with graviton exchange (see e.g. \HijanoZSA). The Regge limit corresponds to  special kinematics, which on the gravity side is described by the scattering of highly energetic
particles whose trajectories in the bulk are approximately null.

Such scattering can be described in the eikonal approximation where particles follow classical trajectories but their wavefunctions
acquire a phase shift $\delta(S,L)$.
The phase shift  is a function of the total energy $S$ and the impact parameter $L$. 
In the CFT language, this phase shift can be extracted from the Fourier transform of the four-point function.
In \CornalbaXM\ the phase shift extracted from the four-point function of the type 
$\langle \OO_1 \OO_1 \OO_2 \OO_2\rangle$ was shown to be equal (up to a factor of $-\pi$) to the anomalous dimension of the double-trace operators $[\OO_{1}\OO_{2}]_{n,l}$
 at leading order in $1/N^2$.
The Regge limit implies that the calculation is valid for   $n,l \gg 1$.
These anomalous dimensions have been subsequently verified in 
\refs{\AldayGDE,\KavirajCXA\VosPQA\AldayEYA\DeyZBG\SleightEPI\AharonyNPF\CardonaDOV\SleightRYU-\CardonaQRT}.

So far both operators $\OO_1$ and $\OO_2$ were assumed to have conformal dimensions of order one.
In the following, we will refer to them as ``light" operators and denote them by $\OO_L$.
In \KulaxiziDXO\  one pair of  operators (which we denote by $\OO_H$) was taken to be ``heavy", with conformal dimension $\Delta_H$ scaling as
the central charge.
The ratio $\mu \sim \Delta_H/C_T$ is a useful expansion parameter;
 its power $k$ corresponds to  the number of stress tensors 
 in the multi-stress tensor operators exchanged in the T-channel ($\OH\times\OH\to (T_{\mu\nu})^k\to \OL\times\OL$)\foot{Recently a similar limit was studied in \FitzpatrickZQZ.}. 
As explained in \KulaxiziDXO, one can define the phase shift as a Fourier transform of the
$\langle \OO_H \OO_H \OO_L \OO_L\rangle$  four-point function. 
It is related to the time delay and angle deflection of a highly energetic particle traveling along a null geodesic
in the background of an asymptotically AdS black hole.
The black hole corresponds to the insertion of the heavy operator $\OH$; its mass in the units of AdS radius is proportional to $\mu$.

The phase shift $\delta(S,L)$ was computed in gravity in \KulaxiziDXO\ as an infinite series expansion in $\mu$, {\it i.e.}, 
%a gravity calculation was performed to compute all coefficients in the expansion 
\eqn\phaseshiftexp{   \delta(S,L) = \sum_{k=1}^\infty \delta^{(k)} \mu^k  \,, }
with terms subleading in $1/N^2$ suppressed.
The anomalous dimensions of heavy-light double-trace operators $[\OH\OL]_{n,l}$ admit a similar expansion
\eqn\anomexp{   \gamma(n,l) = \sum_{k=1}^\infty \gamma^{(k)} \mu^k.}
In \KulaxiziDXO\ it was also proven that  
\eqn\gammaone{ \gamma^{(1)}=-{\delta^{(1)}\over \pi}   } 
where the following identifications are implied:
\eqn\identification{   h=n+l, \qquad \hbar=n, \qquad S=4h\hbar, \qquad  e^{-2L}={\hbar \over h}.}
However, it was observed that this relation does not hold for higher order terms, i.e.\ in general
$\gamma^{(k)}$ is not proportional to $\delta^{(k)}$.
One of the aims  of this paper is to explain how higher order anomalous dimensions are related to higher order terms
in the phase shift.

%Note that in $d=2$ a similar correlator has been studied in greater detail using the full Virasoro symmetry \refs{\FitzpatrickZHA\FitzpatrickDLT\FitzpatrickVUA\CollierEXN-\KulaxiziDXO}. Among other things, one interesting question to ask is how the information loss problem is solved within CFT. This has been studied in two dimensions in e.g.\ \refs{\FitzpatrickMJQ,\FitzpatrickIVE} by analyzing a similar heavy-heavy-light-light 4-pt correlation function. 

\subsec{Summary of the results}

\noindent In this paper we explain how to compute the anomalous dimensions of heavy-light double-trace
operators $[\OH\OL]_{n,l}$ order by order in $\mu$, using the phase shift result of \KulaxiziDXO.
In particular, we show that the  $\OO(\mu^2)$  anomalous dimensions in any $d$ are given by
\eqn\Result{ 
 \gamma^{(2)}=-{\delta^{(2)}\over \pi}+{\gamma^{(1)}\over 2}(\partial_h+\partial_\hbar)\gamma^{(1)}, \qquad \DH\gg l, n \gg 1.
}
 Using known results for $\delta^{(1)}$ and $\delta^{(2)}$ from \KulaxiziDXO, we find an explicit expression for  $\gamma^{(2)}$ 
 and compare it with the known results in the lightcone limit ( $\DH\gg l \gg n \gg 1$).
 We find perfect agreement.

The rest of the paper is organized as follows. In the next section, we review the 4-point function with two heavy scalar operators $\OH$ and two light scalar operators $\OL$. 
This is the main object studied in this paper and we refer to it as a heavy-heavy-light-light correlator. 
We consider holographic CFTs, where  the T-channel exchange (where $\OH \rightarrow \OH$ and $\OL \rightarrow \OL$)
is dominated by multi-trace operators made out of the stress tensor.
We relate this to corrections to the CFT data in the S-channel (OPE coefficients and anomalous dimensions).

In Section 3 we focus on four-dimensional holographic CFTs. At  $\OO(\mu)$, we use the crossing equation between the S- and T-channel to solve for the anomalous dimensions of heavy-light double-trace operators $[\OH\OL]_{n,l}$. The result is eq. \gammaone, valid for $l, n\gg1$.
We then introduce  the impact parameter representation  which allows us to  rewrite the S-channel expansion as a Fourier transform. 
We use this  to relate the phase shift  to the anomalous dimensions of $[\OH\OL]_{n,l}$ at  $\OO(\mu^2)$,
thereby deriving \Result. 
Using a known result for the phase shift $\delta^{(2)}$, we write down an explicit expression for $\gamma^{(2)}$. In the subsequent  $l \gg n$ limit it reduces to 
the result which has been obtained in \KulaxiziDXO\ in a completely different way
(by computing  corrections to the energies of excited states in the AdS-Schwarzschild background).

In Section 4 we generalize these results to any $d$ ($d=2$ is treated separately in Appendix D). 
By solving the Casimir equation in the  limit $\DH\gg \DL,l,n$, we obtain the conformal blocks for heavy-light double-trace operators in the S-channel.
 Using the explicit expression for the blocks together with the mean field theory OPE coefficients,
 we derive an impact parameter representation valid in general dimensions. 
Just as in the $d=4$ case, this allows us to write the S-channel sum as a Fourier transform. 
Hence, we show that \Result\ holds for any $d$. 
We compute $\gamma^{(2)}$ in the lightcone limit and find perfect agreement with the results quoted in \KulaxiziDXO. 
%Also, we have confirmed the validity of the derivative relation for the first correction to the OPE coefficients in the Regge limit. 
In addition, we find an expression for  the $\OO(\mu^2)$ corrections to the OPE coefficients.

Section 5 discusses various observations and mentions some open problems. 
Appendices contain additional technical details.
The conformal bootstrap calculations are summarized in Appendix A, the proof of the impact parameter representation in $d=4$ in Appendix B and the proof in general dimension $d$ in Appendix C. The special case of $d=2$ is treated in Appendix D.
Appendix E discusses the fate of some boundary terms. Appendices F and G contain some identities
which are used in Section 5.

\newsec{Heavy-heavy-light-light correlator in holographic CFTs}
In this section, crossing relations for a heavy-heavy-light-light correlator of pairwise identical scalars are reviewed. We consider large $N$ CFTs, with $N^2\sim\CT$ and $\CT$ the central charge, with a parametrically large gap $\Delta_{\rm gap}$ in the spectrum of single trace operators with spin $J>2$. The object that we study is a four-point correlation function between two light scalar operators $\OL$, with scaling dimension of order one, and two heavy scalar operators $\OH$, with scaling dimension $\DH$ of  $\OO(C_T)$. Explicitly, we  expand the CFT data in the parameter $\mu$ defined in \KulaxiziDXO\ as 
\eqn\MuDef{
\mu={{4\Gamma(d+2)}\over{(d-1)^{2}\Gamma(d/2)^{2}}}{{\Delta_{H}}\over{\CT}}, 
}
which is kept fixed as $C_T\to\infty$. Our conventions mostly follow those of \refs{\KulaxiziDXO}.

The four-point function  is fixed by conformal symmetry up to a function $\AA(u,v)$ of the cross-ratios as
\eqn\fourptfcn{
\langle \OH(x_4)\OL(x_3)\OL(x_2)\OH(x_1)\rangle = {\AA(u,v)\over x_{14}^{2\DH}x_{23}^{2\DL}},
}
where $u,v$ are cross-ratios
\eqn\CrossRatios{\eqalign{
  u &= z\zbar = {x_{12}^2x_{34}^2\over x_{13}^2x_{24}^2}\cr
  v &= (1-z)(1-\zbar) = {x_{14}^2x_{23}^2\over x_{13}^2x_{24}^2}
}}
and $x_{ij}=x_i-x_j$. Using conformal symmetry we can fix $x_1=0$, $x_3=1$ and $x_4\to\infty$, with the last operator confined to a plane parameterized by $(z,\zbar)$. It will also be convenient to introduce the following coordinates after the analytic continuation  ( $z\rightarrow ze^{-2i\pi}$):
\eqn\Coordinates{\eqalign{
	1-z &=\sigma e^{\rho}\cr 
	1-\zbar &=\sigma e^{-\rho}.
}}
The Regge limit then corresponds to $\sigma\to 0$ with $\rho$ kept fixed. 

The main object of study is an appropriately rescaled version of \fourptfcn\ 
\eqn\GDef{
  G(z,\zbar) = \lim_{x_4\to\infty} x_4^{2\DH}\langle\OH(x_4)\OL(1)\OL(z,\zbar)\OH(0)\rangle.
}
This can be expanded in the S-channel $\OL(z,\zbar)\to\OH(0)$ as
\eqn\OPESChannel{
  G(z,\zbar) = (z\zbar)^{-{1\over 2}(\DH+\DL)}\sum_{\OO}\left(-{1\over 2}\right)^J\lambda_{\OH\OL\OO}\lambda_{\OL\OH\OO}g_{\OO}^{\DHL,-\DHL}(z,\zbar),
}
where $\DHL=\DH-\DL$, $\lambda_{ijk}$ are OPE coefficients and the sum runs over primaries $\OO$ with spin $J$ and corresponding conformal blocks $g_\OO$. The correlator can likewise be expanded in the T-channel $\OL(z,\zbar)\to\OL(1)$ as
\eqn\OPETChannel{
  G(z,\zbar) = {1\over [(1-z)(1-\zbar)]^{\DL}}\sum_{\OO'}\left(-{1\over 2}\right)^{J'}\lambda_{\OH\OH\cal \OO'}\lambda_{\OL\OL\OO'} g_{\OO'}^{0,0}(1-z,1-\zbar),
}
where we again sum over primaries $\OO'$ with spin $J'$. The equality of \OPESChannel\ and \OPETChannel\ constitutes an example of a crossing relation, in both channels we sum over an infinite set of conformal blocks $g_{\OO}^{\Delta_1,\Delta_2}(z,\zbar)$. These contain the contribution from a primary $\OO$ and all its descendants. 
(For recent reviews on conformal bootstrap see \refs{\PolandEPD\SimmonsDuffinGJK-\RychkovIQZ}.)
 Here we have distinguished between operators $\OO$ and $\OO'$, in the S- and T-channel, respectively, in order to stress that generically different operators are relevant in different channels. As an example of this, in the lightcone limit in $d>2$ one finds \refs{\FitzpatrickYX,\KomargodskiEK} that the T-channel is dominated by the identity operator, while in the S-channel an infinite number of operators contribute. These are the so-called double-twist operators that exist at large spin in any CFT$_{d>2}$. 

We will assume the following scaling for a non-trivial single trace operator $\OO$ (not including the stress tensor)
\eqn\Scaling{
  \langle \OO_{H,L}\OO_{H,L}\OO\rangle \sim{1\over \sqrt{\CT}}. 
}
The conformal Ward identity fixes the following 3-pt function for the stress tensor
\eqn\Ward{
  \langle \OO_{H,L}\OO_{H,L} T_{\mu\nu}\rangle\sim \Delta_{H,L}, 
}
which implies the following scaling for the exchange of the stress tensor in the T-channel
\eqn\StressTensorExchange{
  {\langle \OH\OH T_{\mu\nu}\rangle\langle T_{\mu\nu}\OL\OL\rangle\over \CT}\sim {\DH\DL\over \CT}\sim \mu. 
}
Keeping $\mu$ small, it follows that the leading contribution in the T-channel is given by the disconnected correlator $\langle\OH\OH\rangle\langle\OL\OL\rangle$, i.e., the exchange of the identity
operator.  Decomposing the disconnected correlator in the S-channel, 
%e.g., using the OPE inversion formula \refs{\CaronHuotVEP,\SimmonsDuffinNUB} or by conglomeration of operators \FitzpatrickDM, 
we will infer the existence of the  ``double-trace operators'' $[\OH\OL]_{n,l}$ for all integers $n,l$, with scaling dimension $\Delta=\Delta_H+\Delta_L+2n+l+\gamma(n,l)$ and spin $l$.
 Moreover, the OPE coefficients scale as (the explicit expression is given below)
\eqn\ThreeptScaling{
  \langle\OH\OL[\OH\OL]_{n,l}\rangle\sim 1.
}

Keeping $\mu\sim \DH/\CT$ fixed as $\CT\to\infty$, \StressTensorExchange\ implies that the CFT data of double-trace operators $[\OH\OL]_{n,l}$ receives perturbative corrections
 in $\mu$. 
 We therefore expand the anomalous dimensions of these double-trace operators, as well as the OPE coefficients 
 \eqn\opecoeff{  P_{n,l}\equiv \left( -{1\over2}\right)^l \, \lambda_{\OH\OL[\OH\OL]_{n,l}}\lambda_{\OL\OH[\OH\OL]_{n,l}}  \,, }
 in $\mu$ as
\eqn\MuExpansion{\eqalign{
  &\gamma(n,l) = \mu\gamma^{(1)}+\mu^2\gamma^{(2)}+\ldots\cr
  &  P_{n,l} = P^{(0)}(1+\mu P^{(1)}+\mu^{2} P^{(2)} \ldots),
}}
with $\ldots$ denoting higher order terms. 

To reach the Regge limit we analytically continue $z\to \ee^{-2\pi i}z$, under which the blocks in the S-channel transform as (see e.g. \refs{\MaldacenaIUA,\ \LiLMH})
\eqn\BlocksAnalyticContinuation{
  g_{\Delta,J}(z,\zbar)\to \ee^{-i\pi(\Delta-J)}g_{\Delta,J}(z,\zbar).
}
In particular, for double-trace operators $[\OH\OL]_{n,l}$ with scaling dimension $\Delta=\DH+\DL+2n+l+\gamma(n,l)$, the blocks transform as
\eqn\DTOAnalyticContinuation{
  g_{[\OH\OL]_{n,l}}^{\DHL,-\DHL}(z,\zbar) \to e^{-i\pi(\DH+\DL)}e^{-i\pi\gamma(n,l)}g_{[\OH\OL]_{n,l}}^{\DHL,-\DHL}(z,\zbar).
}
In what follows it will be convienent to do a change of variables to $h=n+l$ and $\hbar=n$ and to denote the block due to a heavy-light double-trace operator $[\OH\OL]_{\hbar,h-\hbar}$ as $g_{h,\hbar}^{\DHL,-\DHL}$. Substituting the $\mu$ expansion \MuExpansion\ in the S-channel \OPESChannel\ and performing the usual analytic continuation to $\OO(\mu)$ leads to
\eqn\OPESAnalyticallyContinued{\eqalign{
  G(z,\zbar)|_{\mu^0}=(z\zbar)^{-{1\over 2}(\DH+\DL)}\sum_{h\geq\hbar\geq0}^\infty P^{(0)}& g_{h,\hbar}^{\DHL,-\DHL}(z,\zbar)\cr
  G(z,\zbar)|_{\mu^1}=(z\zbar)^{-{1\over 2}(\DH+\DL)}\sum_{h\geq\hbar\geq0}^\infty P^{(0)}&\left(P^{(1)}+\gamma^{(1)}\left({1\over 2}\left(\pa_h+\pa_{\hbar}\right)-i\pi\right)\right)\cr
  &\times g_{h,\hbar}^{\DHL,-\DHL}(z,\zbar).\cr
}}
The new single trace operators that can possibly appear here would be subleading in $1/N^2$.
Continuing to  $\OO(\mu^2)$, the imaginary part of the S-channel is given by

\eqn\OrderMuSqImag{\eqalign{
	{\rm Im}&(G(z,\zbar))|_{\mu^2}=-i\pi (z\zbar)^{-{1\over 2}(\DH+\DL)}\times\cr
	&\times\sum_{h\geq\hbar\geq0}^\infty P^{(0)}\Big(\gamma^{(2)}+\gamma^{(1)}P^{(1)}+{(\gamma^{(1)})^2\over 2}(\pa_h+\pa_\hbar)\Big) g_{h,\hbar}^{\DHL,-\DHL}(z,\zbar).
}}

\noindent Moreover, the real part of the correlator at the same order is given by
\eqn\musqreal{\eqalign{{\rm Re}&(G(z,\zbar))|_{\mu^2}=(z\bar{z})^{-{1\over 2}(\DH+\DL)}\sum_{h\geq\hbar\geq0}^\infty P^{(0)}\Big(P^{(2)}-{1\over 2}\pi^{2}(\gamma^{(1)})^{2}+\cr
&+{1\over 2}(\gamma^{(2)}+P^{(1)}\gamma^{(1)}) (\partial_{h}+\partial_{\bar{h}})+{1\over 8}(\gamma^{(1)})^{2}(\partial_{h}+\partial_{\bar{h}})^{2} \Big)g_{h,\hbar}^{\DHL,-\DHL}(z,\zbar)\,.
}}

The product of OPE coefficients $P^{(0)}$ is fixed by the correlator at  $\OO(\mu^{0})$ in \OPESAnalyticallyContinued\ and can be found in \FitzpatrickDM:
\eqn\OpeCoeff{\eqalign{
  P^{(0)} = &{(\DH+1-d/2)_{\hbar}(\DL+1-d/2)_{\hbar}(\DH)_h(\DL)_h\over \hbar!(h-\hbar)!(\DH+\DL+\hbar+1-d)_{\hbar}(\DH+\DL+h+\hbar-1)_{h-\hbar}}\cr 
  &\times {1\over (h-\hbar+d/2)_{\hbar}(\DH+\DL+h-d/2)_{\hbar}},
}}
where $(a)_b$ is the Pochhammer symbol. In the limit $\DH\gg \DL,h,\hbar$, \OpeCoeff\ simplifies
\eqn\OpeCoeffHeavy{
  P^{(0)}\approx C_{\DL}{\Gamma(\DL+\hbar-d/2+1)\Gamma(\DL+h)\over \hbar!(h-\hbar)!(h-\hbar+d/2)_\hbar},
}
where $C_{\DL}=(\Gamma(\DL-d/2+1)\Gamma(\DL))^{-1}$. 
As we will see below, in the Regge limit the dominant contribution in the S-channel comes from double-trace operators with $h,\hbar\gg 1$.
 In this limit the OPE coefficients are given by
\eqn\ReggeOpe{
	P^{(0)}\approx C_{\DL}(h\hbar)^{\DL-{d\over 2}}(h-\hbar)^{{d\over 2}-1}.
}
We will further need $\lambda_{\OL\OL T}\lambda_{\OH\OH T}$ in \OPETChannel, these are fixed by Ward Identities to be 
\eqn\OpeStressTensor{
  {\lambda_{\OL\OL T}\lambda_{\OH\OH T}\over \DL} =\left({d\over d-1}\right)^2{\DH\over \CT} =  {\mu\Gamma\left({d\over 2}+1\right)^{2} \over \Gamma\left(d+2\right)}.
}
Note that as explained in \KulaxiziDXO, an expansion in $\mu$ corresponds in the bulk to an expansion in the black hole Schwarzschild radius in AdS units. 

\newsec{Anomalous dimensions of heavy-light double-trace operators in $d=4$}
In this section we investigate the anomalous dimensions of  heavy-light double-trace operators $[\OH\OL]_{\hbar,h-\hbar}$ in $d=4$ using conformal bootstrap.
%similar to method of \FitzpatrickYX. 
Moreover, using a four-dimensional impact parameter representation we relate the anomalous dimensions to the bulk phase shift to  $\OO(\mu^2)$.
This procedure can be repeated order by order in $\mu$ to obtain the OPE data (anomalous dimensions and OPE coefficients -- see also Section 4) to the desired order.

\subsec{Anomalous dimensions in the Regge limit using bootstrap}
The conformal blocks in $d=4$ are given by \DolanUT
\eqn\BlocksFourD{
  g_{\Delta,J}^{\Delta_{12},\Delta_{34}}(z,\zbar) = {z\zbar\over z-\zbar}\left(k_{\Delta+J}(z)k_{\Delta-J-2}(\zbar)-(z\lra\zbar)\right)
}
where
\eqn\KFunction{
  k_{\beta}(z) = z^{\beta/2}{}_2F_1\left({\beta-\Delta_{12}\over 2},{\beta+\Delta_{34}\over 2},\beta,z\right).
}
In the limit $\DH\sim\CT \gg 1$ the hypergeometric functions  in \BlocksFourD\ can be substituted by the identity up to $1/\DH$ corrections. 
Explicitly, the conformal blocks of $[\OH\OL]_{\hbar,h-\hbar}$ in the heavy limit are given by
\eqn\BlockDTOFourD{
  g_{h,\hbar}^{\DHL,-\DHL}(z,\zbar)= {(z\zbar)^{{1\over 2}(\DH+\DL)}(z^{h+1}\zbar^{\hbar}-z^{\hbar}\zbar^{h+1})\over z-\zbar}.
}
Inserting this form of the conformal blocks in \OPESAnalyticallyContinued\ together with the OPE coefficients in the Regge limit \ReggeOpe, we approximate the sums by integrals and find the following expression at  $\OO(\mu^{0})$ in the S-channel 
\eqn\Leading{
  G(z,\zbar)|_{\mu^0}= {C_\DL\over z-\zbar}\int_0^\infty dh\int_0^hd\hbar (h\hbar)^{\DL-2}(h-\hbar)\left(z^{h+1}\zbar^\hbar-z^\hbar\zbar^{h+1}\right).
}
The integrals are explicitly computed in Appendix A; the result is the disconnected correlator in the T-channel $[(1-z)(1-\zbar)]^{-\DL}$ in the Regge limit $\sigma\to0$.

At  $\OO(\mu)$ in holographic CFTs the leading corrections in the T-channel come from the exchanges of the stress tensor and double-trace operators $[\OL\OL]_{n,l=2}$
($[\OH\OH]_{n,l=2}$  are heavy and therefore decouple). 
The conformal block for the T-channel exchange of the stress tensor is found after $z\to e^{-2\pi i}z$ to be given by 
\eqn\StressTensorBlockFourD{
	g_{T_{\mu\nu}} = {360i\pi e^{-\rho}\over \sigma (e^{2\rho}-1)}+\ldots,
} 
where $\ldots$ denotes non-singular terms. The contribution from the stress tensor exchange in the T-channel is thus imaginary for real values of $\sigma$ and $\rho$.
 The only imaginary term at order $\mu$ in the S-channel expansion \OPESAnalyticallyContinued\ comes from the term proportional to $-i\pi\gamma$;
it must  reproduce  \StressTensorBlockFourD. 
%Here we assume that the OPE coefficients and the anomalous dimensions of heavy-light double-trace operators in \OPESAnalyticallyContinued\ are real. 

%Noting that the only imaginary term in \OPESAnalyticallyContinued\ in the S-channel comes from the term proportional to $-i\pi\mu\gamma^{(1)}$, this contribution has to contain the stress tensor exchange in the T-channel. 
%\eqn\BootStrapFirstOrderFourD{\eqalign{
%	{1\over \sigma^{2\DL}}{360i\pi\lambda_{\OH\OH T_{\mu\nu}}\lambda_{\OL\OL T_{\mu\nu}}e^{\rho}\over \sigma(e^{2\rho}-1)} &= i\pi\mu\ee^{-i\pi(\DH+\DL)}{e^{-\rho}\over \sigma (e^{2\rho}-1)}\sum_{h\geq\hbar\geq0}^\infty P^{(0)}\gamma^{(1)}\cr 
%  &\times
%  \left((1-\sigma e^{\rho})^{h+1}(1-\sigma e^{-\rho}))^{\hbar}-(1-\sigma e^{\rho})^{\hbar}(1-\sigma e^{-\rho})^{h+1}\right).
%}}

In the Regge limit, we approximate the sum in the S-channel by an integral and insert the OPE coefficients from \ReggeOpe; the imaginary part at  $\OO(\mu)$ in the S-channel is thus given by
\eqn\ReggeBoot{
  {\rm Im}(G(z,\zbar))|_{\mu^1}={-i\pi C_{\DL}\over z-\zbar}\int_0^\infty dh \int_0^h d\hbar (h\hbar)^{\DL-2}(h-\hbar)\gamma^{(1)}(h,\hbar)\left(z^{h+1}\zbar^\hbar-z^\hbar\zbar^{h+1}\right).
}
With the ansatz $\gamma^{(1)}(h,\hbar)=c_1h^a\hbar^b/(h-\hbar)$ the integrals in \ReggeBoot\ can be computed (for more details see Appendix A). 
In order to reproduce the exchange of the stress tensor, the anomalous dimensions at  $\OO(\mu)$ must be equal to  
\eqn\AnDimFirstOrderFourD{\eqalign{
  \gamma^{(1)} &= -{90\lambda_{\OH\OH T_{\mu\nu}}\lambda_{\OL\OL T_{\mu\nu}}\over \mu\DL}{\hbar^2\over h-\hbar}\cr 
               &= -{3\hbar^2\over h-\hbar},
}}
where in the second line we  inserted the OPE coefficients from \OpeStressTensor. With the form \AnDimFirstOrderFourD\ not only the stress tensor exchange is reproduced, but also an infinite sum of spin-2 double-trace operators $[\OL\OL]_{n,l=2}$ with scaling dimension $\Delta_n= 2\DL+2+2n$.
This is similar to what happens in the light-light case \LiLMH.

To determine the second order corrections to the anomalous dimensions we use the derivative relationship:
\eqn\DerivativeRelation{
  P^{(0)}P^{(1)}={1\over 2} (\pa_h+\pa_\hbar)\left(P^{(0)}\gamma^{(1)}\right).
}
We will prove below (see Section 4.3)  that this relationship is true in the  limit $h, \bar h \gg1$. The imaginary part at  $\OO(\mu^2)$ in the S-channel from \OPESAnalyticallyContinued\ is then given by 
\eqn\SecondOrderFourD{
  {\rm Im}(G(z,\zbar))|_{\mu^2}=-i\pi\int_0^\infty dh\int_0^hd\hbar P^{(0)}\left(\gamma^{(2)}+\gamma^{(1)}P^{(1)}+{(\gamma^{(1)})^2\over 2}(\pa_h+\pa_\hbar)\right)g_{h,\hbar}.
}
With the help of \DerivativeRelation, one can write  \SecondOrderFourD\  as 
\eqn\SecondOrderFourD{\eqalign{
  {\rm Im}(G(z,\zbar))|_{\mu^2}=&-i\pi\int_0^\infty dh\int_0^hd\hbar P^{(0)}\left(\gamma^{(2)}-{\gamma^{(1)}\over 2}(\pa_h+\pa_\hbar)\gamma^{(1)}\right)g_{h,\hbar}\cr
  &+{\rm total\, derivative},
}}
where the total derivate term does not contribute  (see Appendix E for details). In order to fix $\gamma^{(2)}$ completely from crossing symmetry, we would need to consider the exchange of infinitely many double-trace operators made out of the stress tensor in the T-channel. Instead, we will use an impact parameter representation to relate $\gamma^{(2)}$ to the bulk phase shift calculated from the gravity dual in \KulaxiziDXO.

\subsec{$4d$ impact parameter representation and relation to bulk phase shift}
In \CornalbaXM\ the anomalous dimensions of light-light double-trace operators in the  limit $h, \bar h\gg 1$ were shown to be related to the bulk phase shift. An impact parameter representation for the case when one of the operators is heavy was introduced in \KulaxiziDXO, where it was also shown that the bulk phase shift and the anomalous dimensions are equal at  $\OO(\mu)$. The goal of this section is to see explicitly how the bulk phase shift and the anomalous dimensions are related to  $\OO(\mu^2)$. 

The correlator \GDef\ can be written in an impact parameter representation as
\eqn\Impact{
  G(z,\zbar)=\int_0^\infty dh\int_0^hd\hbar\,\II_{h,\hbar}f(h,\hbar),
}
with $\II_{h,\hbar}$ given by
\eqn\IIhhDef{
  \II_{h,\hbar}=(z\zbar)^{-{(\Delta_{H}+\Delta_{L})\over 2}}P^{(0)}g_{h,\hbar}^{\DHL,-\DHL}(z,\zbar)
}
and $f(h,\hbar)$ some function that generically depends on the anomalous dimension and corrections to the OPE coefficients. In particular, for $f(h,\hbar)=1$, \Impact\ is equal to the disconnected correlator. In Appendix B it is shown that $\II_{h,\hbar}$ can be equivalently written as
\eqn\IDefinitionFourD{
	\II_{h,\hbar} \equiv C(\DL)\int_{M^+} {d^4p\over (2\pi)^4} (-p^2)^{\DL-2} e^{-ipx} (h-\hbar)\delta(p\cdot \ebar+h+\hbar)\,\delta\left({p^2\over 4}+h \hbar\right)\,
}
where $M^+$ is the upper Milne wedge with $\{p^2\leq 0, \,\, p^0\geq 0\}$, $C(\DL)$ given by (with $d=4$)
\eqn\Cdelta{
C(\Delta)\equiv {2^{d+1-2\Delta}\pi^{1+{d\over 2}}\over \Gamma(\Delta)\Gamma(\Delta-{d\over 2}+1)}\,
}
and $\ebar = (1,0,0,0)$. Moreover, following \KulaxiziDXO, we will set $z=e^{ix^+}$ and $\zbar=e^{ix^-}$, with $x^+=t+r$ and $x^-=t-r$ in spherical coordinates. 

Using the identity 
\eqn\DeltaIdentity{
	\delta(p\cdot \ebar+h+\hbar)\,\delta\left({p^2\over 4}+h \hbar\right)= {1\over |h-\hbar|}\left(\delta\left({p^+\over 2}-h\right)\delta\left({p^-\over 2}-\hbar\right)+(h\lra\hbar)\right),
}
with $p^+=p^t+p^r$, $p^-=p^t-p^r$, the integrals over $h,\hbar$ in \Impact\ are easily computed. With the identification $h={{p^+}\over{2}}$ and $\hbar={{p^-}\over {2}}$ it follows that a generic term like \Impact\ can be written as a Fourier transform 
\eqn\ImpactFourier{
	\int_0^\infty dh\int_0^hd\hbar\,\II_{h,\hbar}f(h,\hbar)= C(\DL)\int_{M^+} {d^4p\over (2\pi)^{4}} (-p^2)^{\DL-2} e^{-ipx}f\left({p^+\over 2},{p^-\over 2}\right).
}
We thus see that the impact parameter representation allows us to  rewrite the S-channel expression as a Fourier transform. 

The phase shift $\delta(p)$ for a pair of operators $\OH$ and $\OL$, with scaling dimensions $\DH/\CT\propto\mu$ and $\DL/C_T\ll 1$, respectively, was defined in \KulaxiziDXO\ by

\eqn\PhaseShiftDefinition{
	\BB(p) \equiv \int d^4x e^{ipx} G(x)= \BB_0(p)e^{i\delta(p)},
}
where $G(x)$ is given in \GDef\ and $\BB_0(p)$ denotes the Fourier transform of the disconnected correlator. As the OPE data, the phase shift admits an expansion in $\mu$: 
\eqn\PhaseShiftExpansion{
	\delta(p)=\mu\delta^{(1)}(p)+\mu^2\delta^{(2)}(p)+\ldots,
}
where $\ldots$ denotes higher order terms in the expansion. Expanding the exponential in \PhaseShiftDefinition\ in $\mu$ we get 
\eqn\FTExpansion{
	\BB(p) = \BB_0(p)\Big(1+i\mu\delta^{(1)}+\mu^2(-{(\delta^{(1)})^2\over 2}+i\delta^{(2)})+\ldots \Big).
}
With \FTExpansion\ the relationship between the anomalous dimensions and the bulk phase shift to  $\OO(\mu^2)$ 
can be established using \OPESAnalyticallyContinued, \OrderMuSqImag\ and \ImpactFourier:
\eqn\Relation{\eqalign{
	\gamma^{(1)}&=-{\delta^{(1)}\over \pi}\cr
	\gamma^{(2)}&=-{\delta^{(2)}\over \pi}+{\gamma^{(1)}\over 2}(\pa_h+\pa_\hbar)\gamma^{(1)}(h,\hbar).
}}

The phase shift was calculated in closed form to all orders in $\mu$ for the four-dimensional case \KulaxiziDXO, with the first and second order terms given by 
\eqn\PhaseShift{\eqalign{
	\delta^{(1)}&= {3\pi\over 2}\sqrt{-p^2}{e^{-L}\over e^{2L}-1}\cr
	\delta^{(2)}&= {35\pi\over 8}\sqrt{-p^2}{2e^{L}-e^{-L}\over (e^{2L}-1)^3},
}}
where  
\eqn\PandRho{
	-p^2=p^+p^-,~~~~~~~~~\cosh L = {p^++p^-\over 2\sqrt{-p^2}}.
}
Using \PhaseShift\ and \PandRho, the $\OO(\mu)$ corrections to the anomalous dimensions are given by $\gamma^{(1)}=-3n^2/l$, which agrees with \AnDimFirstOrderFourD. From \PhaseShift\ and \Relation, we deduce the anomalous dimensions at $\OO(\mu^2)$:
\eqn\GammaSecondOrderResult{
	\gamma^{(2)}=-{35\over 4}{(2l+n)n^3\over l^3}+9{n^3\over l^2}. 
}
Taking the lightcone limit ($l\gg n \gg 1$) in \GammaSecondOrderResult\ we find 
\eqn\GammaSecondLightCone{
	\gamma^{(2)}_{\rm l.c.} = -{17\over 2}{n^3\over l^2}.
}
The anomalous dimensions in the lightcone limit
\GammaSecondLightCone\ agree with eq.\ (6.40) in \KulaxiziDXO, which 
was obtained independently by considering  corrections to the energy levels in the AdS-Schwarzschild background.

\newsec{OPE data of heavy-light double-trace operators in generic $d$}
In this section we will write the general form of conformal blocks for heavy-light double-trace operators in the  limit $\Delta_H\sim C_{T}\gg1$ and general $d>2$. These blocks will be used to confirm the validity of the impact parameter representation in Appendix C. Using the impact parameter representation the OPE data will be related to the bulk phase shift. 
In particular, we show that \Relation\ remains valid in any number of dimensions and
 find explicit expressions for the corrections to the OPE coefficients up to  $\OO(\mu^{2})$.

\subsec{Conformal blocks in the heavy limit}
In order to find conformal blocks in general spacetime dimension $d$ in the  limit $\Delta_{H}\gg\Delta_{L},h,\bar{h}$, we write them in the following form:
\eqn\notAnsatz{g_{h,\hbar}^{\DHL,-\DHL}(z,\zbar)=(z\bar{z})^{{\DH+\DL}\over{2}}F(z,\bar{z})
,}

\noindent where the function $F(z,\bar{z})$ does not depend on $\Delta_{H}$ and  is symmetric with respect to the exchange $z\leftrightarrow \bar{z}$.
Let us now insert the expression \notAnsatz\ into the Casimir equation 
and consider the leading  $\OO(\Delta_{H})$ term:
\eqn\Casimir{z {{\partial}\over{\partial z}}F(z,\bar{z})+\bar{z} {{\partial}\over{\partial \bar{z}}}F(z,\bar{z})-(h+\bar{h})F(z,\bar{z})=0.
}

\noindent The most general solution to eq. \Casimir\ is:
\eqn\SolutionCas{F(z,\bar{z})=z^{h+\bar{h}} f\Big({{\bar{z}}\over{z}}\Big),
}

\noindent where $f$ is an arbitrary function that satisfies $f({1\over {x}})=x^{-h-\bar{h}}f(x)$, since conformal blocks must be symmetric with respect to the exchange $z\leftrightarrow \bar{z}$. 

The behaviour of the conformal blocks as $z,\bar{z}\to 0$ and $z/\bar{z}$  fixed is given by \refs{\DolanUT, \CostaSCB} 
\eqn\BC{g_{\Delta, l}^{\Delta_{12},\Delta_{34}}(z,\zbar)\to {{l !}\over{({{d}\over{2}}-1)_{l}}}(z\bar{z})^{\Delta\over 2}C_{l}^{({{d}\over{2}}-1)}\Big( {{z+\bar{z}}\over{2\sqrt{z \bar{z}}}}\Big)
,}

\noindent where $\Delta=\Delta_{1}+\Delta_{2}+2n+l$ and $C_{q}^{(p)}(x)$ are the Gegenbauer polynomials. 
%Since we consider the regime where $\Delta_{H}$ is parametrically larger than all other parameters, this form of conformal blocks is valid for all $z$ and $\bar{z}$ as long as $\Delta_{H}$ is large enough. 
Using \BC , we can completely determine the function $f$:
\eqn\fixingf{
f\Big({\bar{z} \over z}\Big)={{(h-\bar{h})!}\over{({{d}\over{2}}-1)_{h-\bar{h}}}} \Big({\bar{z} \over z}\Big)^{{h+\bar{h}}\over{2}}C_{h-\bar{h}}^{({{d}\over{2}}-1)}\Big( {{z+\bar{z}}\over{2\sqrt{z \bar{z}}}}\Big).
}
That is, the conformal blocks in the limit of large $\Delta_H$ are given by

\eqn\CBAD{g_{h,\hbar}^{\DHL,-\DHL}(z,\zbar)={{(h-\bar{h})!}\over{({{d}\over{2}}-1)_{h-\bar{h}}}}(z\bar{z})^{{\Delta_{H}+\Delta_{L}+h+\bar{h}}\over 2}C_{h-\bar{h}}^{({{d}\over{2}}-1)}\Big( {{z+\bar{z}}\over{2\sqrt{z \bar{z}}}}\Big).
}

\noindent It is easy to explicitly check that this form of the conformal blocks agrees with the one we used in $d=4$ in the previous Section.

\subsec{Anomalous dimensions}

In Appendix C we  prove the validity of the impact parameter representation in any $d$.
This means that the derivation of \Relation\  goes through for arbitrary $d$.
%relations between the bulk phase shift and the anomalous dimension of heavy-light double-trace operators in the Regge limit hold for general $d$:
%\eqn\FinalRelation{\eqalign{
 %   \gamma^{(1)}(h,\hbar) &= -{\delta^{(1)}(h,\hbar)\over \pi} ,\cr
 %   \gamma^{(2)}(h,\hbar) &= -{\delta^{(2)}(h,\hbar)\over \pi}+{\gamma^{(1)}\over 2}(\pa_h+\pa_\hbar)\gamma^{(1)}(h,\hbar).
%}}
Using known results for the bulk phase shift from \KulaxiziDXO, we thus find
\eqn\AnomDimGenFirst{
  \gamma^{(1)}= -{\hbar^{{d\over 2}}\over h^{{d\over 2}-1}}{\Gamma(d)\over \Gamma({d\over 2})\Gamma({d\over 2}+1)}{}_2F_1({d\over 2}-1,d-1,{d\over 2}+1,{\hbar\over h}).
}
In the lightcone limit ($h=l\gg\hbar=n$) this reduces to
\eqn\AnomDimGenFirstLightCone{
  \gamma^{(1)}_{l.c.}=-{\hbar^{{d\over 2}}\over h^{{d\over 2}-1}}{\Gamma(d)\over \Gamma({d\over 2})\Gamma({d\over 2}+1)}.
}
Similarly, using \Relation\ together with Eq. (2.29) and Eq. (A.5) from \KulaxiziDXO, we find the $\OO(\mu^2)$ corrections to the anomalous dimensions in the  limit $h, \bar h\gg 1$:
\eqn\AnomDimGenDFull{\eqalign{
  \gamma^{(2)}&=-{\delta^{(2)}\over\pi}+{1\over 2}\gamma^{(1)}\left\{ {2\over h+\bar{h}} \gamma^{(1)}-{\Gamma(d)\over\Gamma\left({d\over 2}\right)^2} {\bar{h}^{{d\over 2}-1} h^{{d\over 2}-1} } {(h-\bar{h})^{3-d}\over h+\bar{h}}\right\} =\cr
&=- \left({\hbar^{d-1}\over h^{d-2}}\right)\, {2^{2 d-4} \Gamma\left(d+{1 \over 2}\right)\over \sqrt{\pi}\Gamma(d) } \,\, _2F_1[2d-3,d-2,d,{\hbar\over h}]  \,\, +\cr
&+{\bar{h}^d h^{2-d}\over (h+\bar{h})}\,\,\, {4\Gamma^2(d)\over  d^2\,\Gamma^4\left({d\over 2}\right)}\,\, \,\left(_2F_1[{d\over 2}-1,d-1,{d\over 2}+1,{\hbar\over h}]\right)^2 +\cr
&+{\hbar^{d-1} (h-\hbar)^{3-d}\over h+\hbar}\,\,\,{\Gamma^2(d)\over d\,\Gamma^4\left({d\over 2}\right)}\,\,_2F_1[{d\over 2}-1,d-1,{d\over 2}+1,{\hbar\over h}]
}}
Taking further the lightcone limit ($h\gg\hbar$) we find that
\eqn\AnomDimGenDLightCone{
  \gamma^{(2)}_{l.c.}={\hbar^{d-1}\over h^{d-2}}{2^{2d-4}\over \pi}\left({d\Gamma\left({d+1\over 2}\right)^2\over \Gamma\left({d+2\over 2}\right)^2}-{\sqrt{\pi}\Gamma\left(d+{1\over 2}\right)\over \Gamma\left(d\right)}\right).
}
The result \AnomDimGenDLightCone\ agrees with Eq. (6.42) in \KulaxiziDXO\ which was obtained independently using  perturbation theory in the bulk. In order to see this explicitly, one should notice the following expression for the hypergeometric function:

\eqn\hyperg{{}_{3}F_{2}(1,-{d\over{2}},-{d\over{2}};1+{d\over{2}},1+{d\over{2}};1)={1\over{2}}\Big(1+{{\Gamma^{4}(1+{d\over{2}})\Gamma(2d+1)}\over{\Gamma^{4}(d+1)}}\Big).
}

\subsec{Corrections to the OPE coefficients}

So far, we have only considered the imaginary part of the S-channel. The real part  at  $\OO(\mu)$ is given by the following expression:
\eqn\RealPart{{\rm Re}(G(z,\zbar))|_{\mu}=(z\bar{z})^{-{1\over 2}(\Delta_H + \Delta_L)}\int_{0}^{+\infty}dh\int_{0}^{h}d\bar{h} P^{(0)}\Big( P^{(1)}+{1\over 2}\gamma^{(1)}(\partial_{h}+\partial_{\bar{h}})\Big) g_{h,\bar{h}}^{\scriptscriptstyle {\Delta_{HL},-\Delta_{HL}}} (z,\bar{z}),
}
which can be rewritten as:
\eqn\RealPartt{\eqalign{{\rm Re}(G(z,\zbar))|_{\mu}&=(z\bar{z})^{-{1\over 2}(\Delta_H + \Delta_L)}   \int_{0}^{+\infty}dh\int_{0}^{h}d\bar{h} g_{h,\bar{h}}^{\scriptscriptstyle {\Delta_{HL},-\Delta_{HL}}}\times \cr
&\times\Big(P^{(0)} P^{(1)}-{1\over 2} (\partial_{h}+\partial_{\bar{h}})(P^{(0)} \gamma^{(1)}) \Big)+{\rm total\, derivative}.
}}

\noindent The total derivative term in \RealPartt\ can be shown to vanish as explained in Appendix E.

To derive a relation  between the corrections to the OPE coefficients and the anomalous dimensions at  $\OO(\mu)$,
let us consider the  limit $h, \bar{h}\gg 1$ and substitute $\bar h$ by $h$ everywhere.
Using  \AnomDimGenFirst ,  one can deduce $\gamma^{(1)}\propto h$. 
Then, it follows that $(\partial_{h}+\partial_{\bar{h}})(P^{(0)} \gamma^{(1)}) \propto P^{(0)}$ and hence the
second term on the right hand side of \RealPartt\ behaves as:
\eqn\RealPartttwo{(z\bar{z})^{-{1\over 2}(\Delta_H + \Delta_L)} \int_{0}^{+\infty}dh\int_{0}^{h}d\bar{h} \Big(-{1\over 2} g_{h,\bar{h}}^{\Delta_{HL},-\Delta_{HL}} (\partial_{h}+\partial_{\bar{h}})(P^{(0)} \gamma^{(1)}) \Big) \propto { 1 \over {\sigma^{2\Delta_{L}}}}.
}

On the other hand, we know that in the Regge limit the leading contribution in the T-channel
at $\OO(\mu)$ comes from the exchange of the stress tensor.
 The real part of its conformal block is proportional to $\sigma^{d}$, so the T-channel result  behaves as $1\over \sigma^{2\Delta_{L}-d}$. 
 This  is way less singular than \RealPartttwo . 
Hence \RealPartttwo\ must be canceled by the first term on the right hand side of \RealPartt, at least in the  limit  $h, \bar h \gg 1$. That is:
%We need the correction to OPE coefficients to cancel all singular terms from \RealPartttwo\ up to the $1\over \sigma^{2\Delta_{L}-d}$. In order to achieve this, we claim that the integrand of \RealPartt\ has to be zero up to the subleading corrections in the Regge limit, that will give a contribution of order $1\over \sigma^{2\Delta_{L}-d}$. So, we conclude that the leading term of the correction to the OPE coefficients is
\eqn\CorrectionsToOPE{P^{(0)}P^{(1)}={1\over 2}(\partial_{h}+\partial_{\bar{h}})(P^{(0)}\gamma^{(1)}).
}
A similar relation holds for the OPE coefficients of light-light double-trace operators, e.g. see \refs{\HeemskerkPN,\FitzpatrickDM, \AldayVKK}. In that case it was observed in \AldayGDE\  that the relation is not exact in $(h,\bar{h})$. We expect the same to be true here.
\noindent Furthermore, the real part at  $\OO(\mu^2)$ was given in \musqreal\ as:
\eqn\musqrealrepeat{\eqalign{{\rm Re}(G(z,\zbar))|_{\mu^2}&=(z\bar{z})^{-{1\over 2}(\DH+\DL)}\sum_{h\geq\hbar\geq0}^\infty P^{(0)}\Big(P^{(2)}-{1\over 2}(\pi\gamma^{(1)})^{2}+\cr
&+{1\over 2}(\gamma^{(2)}+P^{(1)}\gamma^{(1)}) (\partial_{h}+\partial_{\bar{h}})+{1\over 8}(\gamma^{(1)})^{2}(\partial_{h}+\partial_{\bar{h}})^{2} \Big)g_{h,\hbar}^{\DHL,-\DHL}.
}}
Using the impact parameter representation this can be expressed as:
\eqn\OPeSecondOrder{\eqalign{
	{\rm Re}&(G(z,\zbar))|_{\mu^2}= \int_0^\infty dh\int_0^h d\hbar \II_{h,\hbar}\,\,\Big(P^{(2)}-{\pi^2\over 2}(\gamma^{(1)})^2\cr
	&-{1\over 2P^{(0)}}(\partial_{h}+\partial_{\hbar})(P^{(0)}(\gamma^{(2)}+P^{(1)}\gamma^{(1)}))
	+{1\over 8P^{(0)}}(\partial_h+\partial_\hbar)^2(P^{(0)}(\gamma^{(1)})^2)\Big),
}}
where we repeatedly integrated by parts. It follows from \FTExpansion\ and \ImpactFourier, together with $\pi\gamma^{(1)}=-\delta^{(1)}$, that  
the corrections to the OPE coefficients at  $\OO(\mu^2)$ satisfy the following relationship: 
\eqn\OPECoeffSecondOrderP{
  P^{(0)}P^{(2)}={1\over 2}(\partial_{h}+\partial_{\hbar})(P^{(0)}(\gamma^{(2)}+P^{(1)}\gamma^{(1)}))-{1\over 8}(\partial_h+\partial_\hbar)^2(P^{(0)}(\gamma^{(1)})^2).
}

\noindent The arguments above are similar to the ones used in  \refs{\LiLMH , \CornalbaXM}.

\subsec{Flat space limit}

In the flat space limit the relation between the scattering phase shift and the anomalous dimensions 
has been previously discussed in \PaulosSMB.
Hence, it is interesting to consider the flat space limit of eq.  \Result.
 This limit is achieved by taking  the apparent impact parameter to be much smaller than 
the AdS radius.
This corresponds to the small $L$ regime or, equivalently, using $e^{-2 L} =\bar h/h$
to the  $1\ll l \ll n \ll \Delta_{H}$ limit. 

In this limit, according to \AnomDimGenFirst,  the behavior of $\gamma^{(1)}$ is given by
\eqn\gamaone{\gamma^{(1)}\propto n \left({n\over l}\right)^{d-3}.
} 
\noindent Hence, the $\gamma^{(1)}(\partial_{h}+\partial_{\hbar})\gamma^{(1)}$ term in eq. \Result\ behaves as 
\eqn\beh{\gamma^{(1)}\partial_{n} \gamma^{(1)}\propto n \left({n\over l}\right)^{2d-6}.
} 
\noindent Similarly, using equation (A.5) from \KulaxiziDXO, one finds that  $\delta^{(2)}$  behaves as
\eqn\deltabeh{\delta^{(2)}\propto n \left({n\over l}\right)^{2d-5}.
}
\noindent 
Since \beh\ is subleading to \deltabeh,    in the flat space limit the anomalous dimensions
are proportional to the phase shift,  
\eqn\fin{\gamma^{(2)}\approx - {\delta^{(2)}\over \pi}
}

\newsec{Discussion}

\noindent In this paper we studied a four-point function of pairwise identical scalar operators, $\OH$ and $\OL$, in holographic CFTs of any dimensionality. Scaling $\DH$ with the central charge, the CFT data admits an expansion in the ratio $\mu\sim \DH/\CT$ which we keep fixed. 
%Going to the Regge limit, highest spin operators dominate in the channel $\OH\times\OH\to\OO\to\OL\times\OL$.
 Using crossing symmetry and the bulk phase shift calculated in \KulaxiziDXO, we studied  $\OO(\mu^2)$ corrections to the OPE data of heavy-light double-trace operators $[\OH\OL]_{n,l}$ for large $l$ and $n$. In particular, the relationship between the bulk phase shift and the OPE data of heavy-light double-trace operators is found using an impact parameter representation. Furthermore, this allows us in principle to determine the OPE data of $[\OH\OL]_{n,l}$, for  $l, n\gg1$ to all orders in $\mu$, i.e., to all orders in an expansion in the dual black hole Schwarzschild radius.  

Scaling $\DH$ with the central charge enhances the effect of stress tensor exchanges compared to the 
$1/\CT$ corrections due to the exchange of generic operators. 
%(this is also relies on the assumption that $\Delta_{\rm gap}\to \infty$). 
At  $\OO(\mu^2)$ and higher, we therefore expect multi-stress tensor  operators to contribute. 
The OPE coefficients for such exchanges are not known in general.
They would be needed to determine corrections to the OPE data of heavy-light double-trace operators using purely CFT methods.
In a recent paper \FitzpatrickZQZ\ some of these OPE coefficients have been computed.
In particular, the  OPE coefficients with the multi-stress tensor operators of lowest twist have been argued to be universal (independent of the higher derivative couplings in
the bulk gravitational lagrangian).
It would be interesting  to connect these results to the ones discussed in this paper.

 It is a curious fact that each term in the $\mu$-expansion of the bulk phase shift as computed in gravity in \KulaxiziDXO\ can be expressed as an infinite sum of ``Regge conformal blocks'' corresponding to operators of dimension $\Delta=k(d-2)+2 n+2$ and spin $J=2$. Explicitly,
\eqn\deltaksum{
i\,\delta^{(k)}(S,L)= \,f(k)\,\sum_{n=0}^\infty \lambda_k(n) \,\,\, g_{\scriptscriptstyle{ k(d-2)+2n+2,\,2}}^{\scriptstyle R}(S,L) \,,
}
where the coefficients $(f(k),\lambda_k(n))$ are listed in Appendix F and we set $S\equiv\sqrt{-p^2}$ compared to \KulaxiziDXO. Here $g_{\scriptscriptstyle{\Delta,J}}^{\scriptstyle{R}}(S,L)$ denotes a ``Regge conformal block'', and is equal to the leading behaviour of the analytically continued T-channel conformal block in the Regge limit \refs{\SimmonsDuffinUY, \KulaxiziIXA}
\eqn\ReggeBlock{
 g_{\scriptscriptstyle \Delta,J}^{R}(S,L)=i\,c_{\Delta,J}  \,\, S^{J-1}\,\,\Pi_{\Delta-1,d-1}(L)\,\,
 }
defined in terms of 
 \eqn\Reggelimitpar{1-z={e^L\over S},\quad 1-\bar{z}={e^{-L}\over S}
 }
as $S\rightarrow\infty$ and $L$ fixed. Here $c_{\Delta, J}$ are known coefficients which can be found in Appendix F and $\Pi_{\Delta-1,d-1}(L)$ denotes the $(d-1)$-dimensional hyperbolic space propagator for a massive scalar of mass square $m^2=(\Delta-1)$.

To understand the implications of \deltaksum\ let us focus on $k=2$ and consider large impact parameters, a.k.a. the lightcone limit. In this case, one expects that the dominant contribution to the bulk phase shift comes from the infinite sum of the minimal twist double-trace operators built from the stress tensor, schematically denoted by $T_{\mu\nu}\p_{\mu_1}\cdots\p_{\mu_\ell}T_{\rho\sigma}$. \deltaksum\ implies that this infinite sum gives rise to a contribution which can be interpreted as coming from a {\it single} conformal block of an ``effective'' operator of the same twist $\tau=2(d-2)$, but spin $J=2$. At finite impact parameter, one would then need to add the contributions of an infinite tower of such effective operators of twist $\tau=2(d-2)+2 n$ and spin $J=2$, as expressed by the infinite sum in \deltaksum. From this point of view, the coefficients $\lambda_n$ in \deltaksum\ can be interpreted as ratios of sums of OPE coefficients of double-trace operators. It is clear that this picture appears to hold to all orders in $\left({\Delta_H\over C_T}\right)$ or equivalently, the Schwarzschild radius of the black hole.

It would be interesting to investigate whether Rindler positivity constrains the Regge behaviour of the bulk phase shift to grow at most linearly with 
the energy $S$, similarly to Section 5.2 in \KulaxiziIXA. If this were true, one would perhaps only need to understand the origin of the $\lambda_n$ to compute the bulk phase shift to arbitrary order in $\left({\Delta_H\over C_T}\right)$ purely from CFT techniques.

\bigskip
\bigskip

\noindent {\bf Acknowledgments}: 
We thank S. Frolov, G. S. Ng, R. Pereira and P. Di Vecchia for discussions.
The work of R.K. and A.P. is supported in part by the Irish Research Council Laureate Award. The work of P.T. is supported in part by an Ussher Fellowship Award. This work was also supported in part by the National Science Foundation under Grant No. NSF PHY-1748958. M.K. and A.P.  thank KITP Santa Barbara for hospitality during the completion of this work.

\appendix{A}{Details on the conformal bootstrap}
Below we review some of the details of the confomal bootstrap calculations. Explicitly, we will show that exchanges of heavy-light double-trace operators in the S-channel reproduce the disconnected correlator at $\OO(\mu^0)$ and the stress tensor exchange at $\OO(\mu)$.
\subsec{Solving the crossing equation to $\OO(\mu)$  in $d=4$}

We start with the leading $\OO(\mu^{0})$ term in the S-channel that should reproduce the disconnected propagator in the T-channel. This is given in $d=4$ by
\eqn\SChannelLeadingOrder{G(z,\zbar)|_{\mu^{0}} = {{C_{\Delta_{L}}}\over{z-\zbar}}\int_0^\infty dh\int_0^h d\bar{h}(h\bar{h})^{\Delta_{L}-2}(h-\bar{h})(z^{h+1}\zbar^{\bar{h}}-z^{\bar{h}}\zbar^{h+1}).
}

\noindent Let us look at the following piece of \SChannelLeadingOrder :
\eqn\RewrIntegral{\eqalign{-&\int_0^\infty dh\int_0^h d\bar{h}(h\bar{h})^{\Delta_{L}-2}(h-\bar{h})z^{\bar{h}}\zbar^{h+1} = -\int_0^\infty d\bar{h}\int_{\bar{h}}^\infty dh (h\bar{h})^{\Delta_{L}-2}(h-\bar{h})z^{\bar{h}}\zbar^{h+1}\cr
	&= {{\zbar}\over{z}}\int_0^\infty dh\int_h^\infty d\bar{h} (h\bar{h})^{\Delta_{L}-2}(h-\bar{h})z^{h+1}\zbar^{\bar{h}}
\,.}}
Setting $\zbar/z=1$ to leading order in the Regge limit, we find that the S-channel expression reproduces the disconnected correlator:
\eqn\SChannelLeadingOrderFinal{\eqalign{G(z,\zbar)|_{\mu^{0}} &= {{zC_{\Delta_L}}\over{z-\zbar}}\int_0^\infty dh\int_0^\infty d\bar{h}(h\bar{h})^{\Delta_{L}-2}(h-\bar{h})z^{h}z^{\bar{h}}\cr
	&= {{zC_{\Delta_{L}}}\over{z-\zbar}}{{(\log{\zbar}-\log{z})}\over{(\log{z}\log{\zbar})^{\Delta_{L}}}}\Gamma(\Delta_{L})\Gamma(\Delta_{L}-1)\simeq {1\over (1-z)^{\Delta_L}(1-\zbar)^{\Delta_L}}\,.
}}
Notice that to arrive in the last equality we expanded $(z,\bar{z})$ around unity and substituted $C_{\Delta_L}=(\Gamma(\Delta_{L})\Gamma(\Delta_{L}-1))^{-1}$.

Consider now the imaginary part at $\OO(\mu)$ in the S-channel. For convenience we define 
\eqn\Idefdfour{I^{(d=4)}\equiv{\rm Im}(G(z,\zbar))|_{\mu}\,,}
which is then equal to:
\eqn\Sch{\eqalign{
&I^{(d=4)}={-i\pi C_{\DL}\over\sigma(e^{-\rho}-e^{\rho})}\times\cr
&\times\int_0^\infty dh \int_0^hd\bar{h}(h\bar{h})^{\DL-2}(h-\bar{h})\gamma(h,\bar{h}) \Big((1-\sigma e^\rho)^{h+1}(1-\sigma e^{-\rho})^{\bar{h}}-(h\leftrightarrow \bar{h})\Big)\,.
}}
Notice that we used the variables $(\sigma,\rho)$ defined as $z=1-\sigma e^{\rho}$ and $\bar{z}=1-\sigma e^{-\rho}$.

Consider the following ansatz for $\gamma={ch^a\bar{h}^b\over h-\bar{h}}$, where $(a,b, c)$ are numbers tobe determined by the crossing equation. Substituting into \Sch\ and collecting the leading singularity $\sigma^{-k}$ as $\sigma\to 0$ with $k=2\DL+a+b-1$ leads to
\eqn\ComSch{\eqalign{
  &I^{(d=4)}|_{\sigma^{-k}} = {-ic\pi C_{\DL}\over(e^{-\rho}-e^{\rho})}\Big(\Gamma(\DL+a-1)\Gamma(\DL+b-1)(e^{(b-a)\rho}-e^{(a-b)\rho})+\cr
  &+{\Gamma(2\DL+a+b-2)\over \DL+a-1}e^{-(2\DL+a+b-2)\rho}{}_2F_1(\DL+a-1,2\DL+a+b-2,\DL+a,-e^{-2\rho})\cr
  &-{\Gamma(2\DL+a+b-2)\over \DL+a-1}e^{(2\DL+a+b-2)\rho}{}_2F_1(\DL+a-1,2\DL+a+b-2,\DL+a,-e^{2\rho})\Big).
}}
Note that in order to do these integrals we need $\DL+a>1$ and $\DL+b>1$. Using the following identity of the hypergeometric function
\eqn\HyperIdentity{\eqalign{
  {}_2F_1(a,b,c,x) =& {\Gamma(b-a)\Gamma(c)\over \Gamma(b)\Gamma(c-a)}(-x)^{-a}{}_2F_1(a,a-c+1,a-b+1,{1\over x})\cr
    &+{\Gamma(a-b)\Gamma(c)\over\Gamma(a)\Gamma(c-b)}(-x)^{-b}{}_2F_1(b,b-c+1,-a+b+1,{1\over x}),
}}
the third line in \ComSch\ can be simplified and we are left with
\eqn\ComSchRes{\eqalign{
 &I^{(d=4)}|_{\sigma^{-k}}= {ic\pi C_{\DL}\over(e^{2\rho}-1)}\Big(-\Gamma(\DL+a-1)\Gamma(\DL+b-1)e^{(a-b+1)\rho}\cr
 &+{\Gamma(2\DL+a+b-2)\over\DL+a-1}e^{-(2\DL+a+b-3)\rho}{}_2F_1(\DL+a-1,2\DL+a+b-2,\DL+a,-e^{-2\rho})\cr
 &+{\Gamma(2\DL+a+b-2)\over\DL+b-1}e^{-(2\DL+a+b-3)\rho}{}_2F_1(\DL+b-1,2\DL+a+b-2,\DL+b,-e^{-2\rho})\Big).
}}

On the other hand, the Regge limit in the T-channel is dominated by operators of maximal spin. In a holographic CFT, we have $J=2$. If we further take the lightcone limit, $\rho\gg1$, the dominant contribution is due to the stress tensor exchange and behaves as $\sigma^{-1} e^{-(d-1)\rho}$.
To reproduce this behavior from the S-channel, we must set $a=0$ and $b=2$ and make an appropriate choice for the overall constant $c$.
Substituting the designated values of $(a,b,c)$ revals that the first term in \ComSchRes\ precisely matches the T-channel stress tensor contribution, which in the Regge limit (after analytic continuation) behaves like:
\eqn\Tch{
  g_{\Delta,J}\propto {1\over \sigma^{J-1}}{e^{-(\Delta-3)\rho}\over (e^{2\rho}-1)}+\ldots,
}
with $\Delta=d$ and $J=2$. Furthermore, the remaining terms correspond to the exchange of operators with spin $2$ and dimension $2\DL+2+2n$; these are the double-trace operators $[\OL\OL]_{n,l=2}$.

\subsec{Integrating the S-channel result at $\OO(\mu^2)$ in $d=4$}
Below we describe how to use the results for the anomalous dimensions at $\OO(\mu^2)$ in order to recover the imaginary part of the correlator to the same order.
%The result is naturally identified with conformal blocks in the T-channel.  
Using the obtained expressions for the anomalous dimensions \AnDimFirstOrderFourD\ and \GammaSecondOrderResult, we note that the integrand in \SecondOrderFourD\ can be written as 
\eqn\Integrand{\eqalign{
  P^{(0)}\left(\gamma^{(2)}-{\gamma^{(1)}\over 2}(\pa_h+\pa_\hbar)\gamma^{(1)}\right)&= -{35 \bar{h}^3(2h-\bar{h})\over 4(h-\bar{h})^3}P^{(0)}\cr
  &= -{35 h^{\Delta_L-3}\bar{h}^{\Delta_L+1}\over 2\Gamma(\Delta_L-1)\Gamma(\Delta_L)}\sum_{n=0}^\infty \left({\bar{h}\over h}\right)^{n}(1+{n\over 2}).
}
}
Therefore we see that \SecondOrderFourD\ can be written as an infinite sum of integrals of the same form that appeared at $\OO(\mu)$ in \Sch. It then follows that the full S-channel result can be integrated in order to obtain the correlator in position space. Especially, the lightcone result is obtained by setting $k=0$ in \Integrand\ and taking $\rho\to\infty$ which gives
\eqn\LightConeTChannelMuSq{
  {\rm Im}(G(z,\zbar))|_{\mu^2}= {i35\pi\Delta_L(\Delta_L+1)\over 2(\Delta_L-2)}{e^{-3\rho}\over\sigma^{2\Delta_L+1}(e^{2\rho}-1)}+\ldots,
} 
with $\ldots$ denoting terms that are subleading in the lightcone limit. 
The result \LightConeTChannelMuSq\ 
has a form consistent with the contribution 
of an operator with spin-$2$ and $\Delta=6$. 
The full result (beyond the lightcone limit) further contains an infinite number of operators with spin-$2$ of dimension $\Delta=6+2n$ and $\Delta=2\Delta_L+2n+2$. 

\subsec{Solving the crossing equation to $\OO(\mu)$ in $d=2$}
Here we review the calculations needed for the $d=2$ case explained in Appendix D. To $\OO(\mu^0)$ the S-channel \OPESAnalyticallyContinued\ is given by 
\eqn\BootTwoDDisc{ 
  G(z,\zbar)|_{\mu^0} = {1\over \Gamma(\DL)^2}\int_0^\infty\int_0^h d\hbar (h\hbar)^{\DL-1}(z^h\zbar^\hbar+(z\lra \zbar)).
}
The integrand in \BootTwoDDisc\ is symmetric w.r.t.\ $h\lra\hbar$ and can thus be rewritten as 
\eqn\BootTwoDDisc{ 
  G(z,\zbar)|_{\mu^0} = {1\over \Gamma(\DL)^2}\int_0^\infty\int_0^\infty d\hbar (h\hbar)^{\DL-1}z^h\zbar^\hbar,
}
which can easily be seen to reproduce the disconnected correlator $[(1-z)(1-\zbar)]^{-\DL}$ in the Regge limit. 

As in the previous subsection we proceed to consider the imaginary part of the correlator in the S-channel expansion to $\OO(\mu)$. Using a similar notation,
\eqn\Idefdtwo{I^{(d=2)}\equiv{\rm Im}(G(z,\zbar))|_{\mu}\,,}
combined with the ansatz $\gamma_1(h,\hbar)=c \,h^a\hbar^b$, allows us to write:
\eqn\BootTwoD{\eqalign{
I^{(d=2)} = -{ic\,\pi\over \Gamma(\DL)^2}\int_0^\infty\int_0^h d\hbar (h\hbar)^{\DL-1}h^a\hbar^b(z^h\zbar^\hbar+(z\lra \zbar)).
}}
The integrals in \BootTwoD\ can be easily performed given that $a+\DL>0$ and $b+\DL>0$. Changing variables to $z=1-\sigma e^{\rho}$, $\zbar=1-\sigma e^{-\rho}$ and collecting the most singular term $\sigma^{-k}$, with $k=2\DL+a+b$, leads to
\eqn\BootTwoDOneSigma{\eqalign{
  &I^{(d=2)}|_{\sigma^{-k}} = {ic\pi\over \Gamma(\DL)^2}\Big(\Gamma(a+\DL)\Gamma(b+\DL)(-e^{\rho(b-a)}-e^{\rho(a-b)})\cr
  &+{\Gamma(a+b+2\DL)e^{-\rho(a+b+2\DL)}\over a+\DL}{}_2F_1(a+\DL,a+b+2\DL,1+a+\DL,-e^{-2\rho})\cr
  &+{\Gamma(a+b+2\DL)e^{\rho(a+b+2\DL)}\over a+\DL}{}_2F_1(a+\DL,a+b+2\DL,1+a+\DL,-e^{2\rho})\Big).
}}
Using again \HyperIdentity\ we express \BootTwoDOneSigma\ as follows
\eqn\BootTwoDRes{\eqalign{
 & I^{(d=2)}|_{\sigma^{-k}} = {ic\pi\over \Gamma(\DL)^2}\Big(-\Gamma(a+\DL)\Gamma(b+\DL)e^{\rho(a-b)}\cr
  &+{\Gamma(a+b+2\DL)e^{-\rho(a+b+2\DL)}\over a+\DL}{}_2F_1(a+\DL,a+b+2\DL,1+a+\DL,-e^{-2\rho})\cr
  &-{\Gamma(a+b+2\DL)e^{-(a+b+2\DL)\rho}\over b+\DL}{}_2F_1(b+\DL,a+b+2\DL,1+b+\DL,-e^{-2\rho})\Big).
}}
%(Setting $c=i/\pi$ and $a=b=0$, this is seen to reproduce the disconnected correlator $\sigma^{-2\DL}$ in the Regge limit.)
%The exchange of an operator with dimension $\Delta$ and spin $J$ in the T-channel, in the Regge limit is, up to normalization and subleading terms, given by 
%\eqn\BlocktwoDExpr{
%  g_{\Delta,J}\propto {e^{-(\Delta-1)\rho}\over \sigma^{J-1}}+\ldots.
%}
In matching \BootTwoDRes\ with the T-channel expansion, following the same logic as in the previous subsection we deduce that $a=0$ and $b=1$ and fix $c$. The first line in \BootTwoDRes\ then reproduces the exchange of the stress tensor in the T-channel. The other two lines match the contribution of double-trace operators $[\OL\OL]_{n,l=2}$ with dimension $\Delta=2\DL+2n+2$ and spin $2$ in the T-channel expansion. 

\appendix{B}{Details on the impact parameter representation in $d=4$}
Here we will see how the impact parameter representation in four dimensions leads to the expression for the disconnected correlator in the Regge limit, in terms of the integral over $h,\hbar$. 

The objective of this section is to explicitly see that the disconnected contribution of the correlator in the Regge limit
\eqn\scdis{
{1\over [(1-z)(1-\zbar)]^\Delta}= {1\over\Gamma(\Delta)\Gamma(\Delta-1)}\int_0^\infty dh\int_0^h d\hbar (h \hbar)^{\Delta-2}  {h-\hbar\over z-\zbar}\, (z^{h+1}\zbar^\hbar-z^\hbar\zbar^{h+1})\,,
}
can be equivalently written as
\eqn\ipr{
\int_0^\infty dh\int_0^h d\hbar \,\II_{h,\hbar} \,,
} 
with
\eqn\Ihhb{
\II_{h,\hbar} \equiv C(\Delta)\int_{M^+} {d^4p\over (2\pi)^4} (-p^2)^{\Delta-2} e^{-ipx} (h-\hbar)\delta(p\cdot \ebar+h+\hbar)\,\delta\left({p^2\over 4}+h \hbar\right)\,.
}
where $M^+$ is the upper Milne wedge with $\{p^2\leq 0, \,\, p^0\geq 0\}$ and
\eqn\Cdelta{C(\Delta)\equiv {2^{d+1-2\Delta}\pi^{1+{d\over 2}}\over \Gamma(\Delta)\Gamma(\Delta-{d\over 2}+1)}\,,
}
with $d$ the dimensionality of the spacetime, here $d=4$.

In practice, we need to perform the integral over $p$ in \Ihhb. To do so, we will use spherical polar coordinates and write:
\eqn\ipra{\eqalign{
\II_{h,\hbar} =  {C(\Delta)\over (2\pi)^{3}}&\int_{-\infty}^\infty dp^0 \,\int_0^\infty dp^r\,(p^r)^2\int_{-1}^1 d(\cos{\theta})\, \,(-p^2)^{\Delta-2}\,\theta(p^0)\theta(-p^2) \times\cr
&e^{ip^0 x^0}\, e^{-ir p^r\cos{\theta}} \, \left[\delta\left({p^0+p^r\over 2}-h\right)\,\delta\left({p^0-p^r\over 2}-\hbar\right)+h\leftrightarrow \hbar\right]\, .
}}
The overall factor of $(2\pi)$ is simply the result of the integration with respect to the angular variable $\phi$. Next we perform the integral over $\cos{\theta}$:
\eqn\iprb{
\II_{h,\hbar}={C(\Delta)\over (2\pi)^{3}}\int_{-\infty}^\infty dp^0\,\int_0^\infty dp^r\,(p^r)^2 \,(-p^2)^{\Delta-2} \, e^{ip^0 x^0} \left({e^{-irp^r}-e^{irp^r}\over -i r p^r}\right) \,\theta(p^0)\theta(-p^2)\,(\delta\,\delta),
}
where we set
\eqn\deltafast{(\delta\,\delta)\equiv \delta\left({p^0+p^r\over 2}-h\right)\,\delta\left({p^0-p^r\over 2}-\hbar\right)+h\leftrightarrow \hbar \,.}
Notice that 
\eqn\intpr{\eqalign{
\int_0^\infty dp^r \,{p^r\over i r}\,(-p^2)^{\Delta-2}\,  e^{irp^r} (\delta\,\delta)&-\int_0^\infty dp^r \,{p^r\over i r}\,(-p^2)^{\Delta-2}\, e^{-irp^r}(\delta\,\delta)=\cr
&=\int_{-\infty}^\infty dp^r \,{p^r\over i r}\,(-p^2)^{\Delta-2}\,  e^{irp^r} (\delta\,\delta)\,.
}}
Hence we can write \iprb\ as follows
\eqn\iprc{
\II_{h,\hbar}={C(\Delta)\over (2\pi)^{3}} \int_{-\infty}^\infty {dp^+\,dp^-\over 2}\, {p^+-p^-\over i(x^+-x^-)} \,(-p^2)^{\Delta-2} \, e^{{i\over 2} (p^+x^-+p^-x^+)} \,\theta(p^+)\theta(p^-)\, (\delta\,\delta)\,.
}
Performing the last two integrations is trivial due to the delta-functions. The result is
\eqn\iprd{
\II_{h,\hbar}={1\over \Gamma(\Delta)\Gamma(\Delta-1)}\,{h-\hbar\over i(x^+-x^-)} \,(h \hbar)^{\Delta-2} \, (e^{i h x^+}e^{i\hbar x^-} -e^{i \hbar x^+}e^{ih x^-})\,,
}
which allows us to write \ipr\ as follows:
\eqn\iprf{
\int_0^\infty dh\int_0^h d\hbar \,\II_{h,\hbar}={1\over \Gamma(\Delta)\Gamma(\Delta-1)} \int_0^\infty dh\int_0^h d\hbar {h-\hbar\over i(x^+-x^-)} \,(h \hbar)^{\Delta-2} \, (z^h \zbar^\hbar-z^\hbar \zbar^h)\,.
}
Here we also used the identification $(z=e^{ix^+}, \zbar=e^{ix^-})$.

Observe that \iprf\ is equal to \scdis\ in the Regge limit, where
\eqn\Reggeleading{{z\over z-\zbar}\simeq {1\over i(x^+-x^-)},\quad   {\zbar\over z-\zbar}\simeq  {1\over i(x^+-x^-)}\,.}
However, when considering next order corrections in $(x^+,x^-)$ the impact parameter represention may require corrections. Below we show that these are irrelevant for the questions we are interested in. 

\subsec{Exact Fourier transform}
Here we will compute the Fourier transform for the S-channel expression with the identification $(z=e^{ix^+},\zbar=e^{ix^-})$ and show that the leading order results in the Regge limit given in the previous section do not miss any important contributions.

The generic term in the S-channel which we would like to Fourier transform looks like:
\eqn\Schannelterm{
    \int dh \,d\hbar \,g(x^+,x^-) \tilde{f}(h,\hbar)\,,
}
where 
\eqn\GDefinition{
    g(x^+,x^-)={e^{i(1+h)x^+} e^{i \hbar x^-}-e^{i \hbar x^+} e^{i (h+1) x^-}\over (e^{ix^+}-e^{ix^-})}\,,
}
and 
\eqn\ftilde{
    \tilde{f}(h,\hbar)=i\pi (h \hbar)^{\Delta-2} (h-\hbar) f(h,\hbar)\,,
}
where $f(h,\hbar)$ stands for all the contributions in the S-channel to a given order.

The Fourier transform is:
\eqn\fta{\int d^4x\, e^{i p x} \int dh \,d\hbar \,g(x^+,x^-) \tilde{f}(h,\hbar)=  \int dh \,d\hbar \tilde{f}(h,\hbar) \int d^4x\, e^{i p x} g(x^+,x^-)\,,}
where we simply reversed the order of integration. Our focus in what follows will be the integral:
\eqn\idef{I\equiv \int d^4x\, e^{i p x} g(x^+,x^-)\,.}
Since $x^+=t+r$ and $x^-=t-r$, it is convenient to use spherical polar coordinates to perform the integration. The angular integration over $\phi$ gives us an overall factor of $(2\pi)$ as the integrand is independent of $\phi$. Next we perform the integration over the other angular variable. Similar to what was discussed in the previous section, 
\eqn\thetaint{\int_{-1}^1 d(\cos\theta)\, e^{i p^r r\cos\theta}={e^{ir p^r}-e^{-ir p^r}\over i r p^r}\,.}
Combining the above we can write:
\eqn\Ib{I=2\pi \int_{-\infty}^\infty dte^{-itp^t}\int_0^\infty dr r\,{e^{ir p^r}-e^{-ir p^r}\over i p^r}\, g(t,r)\,.}
It is easy to see that $g(t,r)=g(t,-r)$ and as a result:
\eqn\rint{\int_0^\infty dr\, r e^{-irp^r} \,g(t,r)=-\int_{-\infty}^0 dr\,r e^{irp^r}\, g(t,r)\,,}
which allows us to write the integral as:
\eqn\Ic{I=2\pi \int_{-\infty}^\infty  {dx^+ dx^-\over 2}e^{i p\cdot x} {x^+-x^-\over i (p^+-p^-)} g(x^+,x^-) \,.}
Here $e^{ip\cdot x}=e^{-{i\over 2}(p^+ x^-+p^-x^+)}$ and the above integral can be thought of as a two-dimensional Fourier transform.

To proceed we need the explicit form of $g(x^+,x^-)$ which we write as 
\eqn\gReggeexp{g(x^+,x^-)={e^{ihx^+} e^{i \hbar x^-}\over 1-e^{-i(x^+-x^-)} }+(x^+\leftrightarrow x^-) \,}
and then expand the denominator in the Regge limit
\eqn\denexp{{1\over 1-e^{-i(x^+-x^-)}} ={1\over i(x^+-x^-)}\left[1-{i\over 2} (x^+-x^-)+\cdots\right]\,.}
Substituting into \Ic\ leads to:
\eqn\Id{I=2\pi  {1\over (-p^++p^-)} \int  {dx^+ dx^-\over 2} e^{i p\cdot x} \left\{ e^{ihx^+} e^{i \hbar x^-} \left[1-{i\over 2} (x^+-x^-)+\cdots\right] +(x^+\leftrightarrow x^-) \right\}  \,.  }
Let us compute the integral term by term. The leading term in the Regge limit yields the standard delta functions:
\eqn\Fourierterma{\eqalign{
I_0&=2^{2}\pi^3\, {1\over p^--p^+}\delta({p^+\over 2}-\hbar)\delta({p^-\over 2}-h)+(p^+\leftrightarrow p^-)=\cr
&=2\pi^3\, {1\over h-\hbar}\left\{\delta({p^+\over 2}-\hbar)\delta({p^-\over 2}-h)+(p^+\leftrightarrow p^-) \right\}=\cr
&=2\pi^3\, {1\over h-\hbar} \delta(p\cdot \bar{e}+h+\hbar)\delta({p^2\over 4}+h\hbar)
\,.}}
The subleading terms on the other hand produce the same result except that the delta functions are replaced with derivatives of themselves with respect to $p^r={p^+-p^-\over 2}$.

Let us now consider the full result which up to an overall numerical coefficient can be written as:
\eqn\finalfourier{\int dh\,d\hbar \,\tilde{f}(h,\hbar) \left(1-{\p\over \p p^r}+\cdots \right) \delta(p\cdot \bar{e}+h+\hbar)\delta({p^2\over 4}+h\hbar) \,.}
To evaluate the terms with derivatives of the delta function we need to integrate by parts. Now recall that we are interested in the imaginary piece of the S-channel whose leading behaviour is $\sim\sqrt{-p^2}$  (this dependence is hidden in what we called $\tilde{f}$). It is obvious that the derivatives will produce subleading terms which we are not interested in.

What about the other pieces in the S-channel which are not imaginary?  To $\OO(\mu^2)$ in this case, we know that the leading behaviour grows like $\sim(\sqrt{-p^2})^2$, so by differentiation, a term of the order $\sqrt{-p^2}$ may be produced. However, it is clear that this term will never contribute to the {\it imaginary} term of the S-channel (note that the coefficient in the first term in the parenthesis in \finalfourier\ is real). We thus deduce that the subleading terms in \Id\ are irrelevant for our study. 

\appendix{C}{Impact parameter representation in general spacetime dimension $d$}

Here we want to prove the following equation for general spacetime dimension $d$:

\eqn\iprr{\II_{h,\bar{h}}=(z\bar{z})^{-{(\Delta_{H}+\Delta_{L})\over 2}}P^{(0)}g_{h,\hbar}^{\DHL,-\DHL}(z,\zbar),
}

\noindent using the form of conformal blocks given in \CBAD. We start with the definition of $\II_{h,\bar{h}}$ that is given as:
\eqn\IPRd{\II_{h,\bar{h}}=C(\Delta_{L}) \int_{M^{+}}{{d^{d}p}\over{(2\pi)^{d}}}(-p^{2})^{\Delta_{L}-{{d}\over{2}}}e^{-ipx}(h-\bar{h})\delta(p\cdot\bar{e}+h+\bar{h})\delta({{p^{2}\over{4}}+h\bar{h}}),
}

\noindent where:

\eqn\Cdelta{
C(\Delta_{L})\equiv {2^{d+1-2\Delta_{L}}\pi^{1+{d\over 2}}\over \Gamma(\Delta_{L})\Gamma(\Delta_{L}-{d\over 2}+1)}\, .
}

\noindent Using spherical coordinates we write \IPRd\ as:

\eqn\first{\eqalign{&\II_{h,\bar{h}}=C(\Delta_{L}) \int_{-\infty}^{\infty} dp^{t}e^{ip^{t} t}\int_{0}^{\infty}dp^{r}(p^{r})^{d-2}
\int_{S_{d-2}}  \sin^{d-3}{\phi_{1}}  d\phi_1 \,d\Omega_{d-3}   \cr 
& \times e^{-ip^{r}r\cos{\phi_{1}}}(-p^{2})^{\Delta_{L}-{d\over{2}}}
\theta(-p^{2})\theta(p^{t})\Big\{ \delta\left({{p^{t}+p^{r}}\over{2}}-h\right) \delta\left({{p^{t}-p^{r}}\over{2}}-\bar{h}\right) + (h \leftrightarrow \bar{h}) \Big\}\,,
}}
where $\Omega_{d-3}={2\pi^{d-2\over 2}\over\Gamma\left({d-2\over 2}\right)}$ denotes the area of the unit $(d-3)$-dimensional hypersphere.

\noindent Notice now that
\eqn\Integralfirst{\int_{0}^{\pi}\sin^{d-3}{\phi_{1}}e^{-ip^{r} r\cos{\phi_{1}}}d\phi_{1}=\sqrt{\pi}\Gamma({{d\over{2}}-1})   {}_0F_{1}({{d-1}\over{2}};-{1\over 4}(p^{r})^{2}r^{2})\,.
}
\noindent Substituting \Integralfirst\ back in to \first, one is left with integrals with respect to $p^{t}$ and $p^{r}$ only. These integrals are trivial due to the presence of delta functions.\foot{One only needs to remember that $h\geq \bar{h}\geq 0$.} When these integrations are done, the expression for $\II_{h,\bar{h}}$ is given as:

\eqn\final{\II_{h,\bar{h}}={{2^{3-d} \sqrt{\pi}}\over{\Gamma(\Delta_{L})\Gamma(\Delta_{L}-{d\over{2}}+1)}}e^{it(h+\bar{h})}(h-\bar{h})^{d-2}(h\bar{h})^{\Delta_{L}-{d\over{2}}}{}_0F_{1R}({{d-1}\over{2}};-{1\over 4}(h-\bar{h})^{2}r^{2})
,}
where ${}_0F_{1R}(a,x)=\Gamma(a)^{-1}{}_0F_{1}(a,x)$.
\noindent Relations between coordinates $t$ and $r$ with $x^{+}$ and $x^{-}$ are given as: $x^{+}=t+r$ and $x^{-}=t-r$.

On the other hand, using the explicit form for conformal blocks \CBAD\ and OPE coefficients in the Regge limit \ReggeOpe\ one finds that:
\eqn\aa{\eqalign{(z\bar{z})^{-{(\Delta_{H}+\Delta_{L})\over 2}}&P^{(0)}g_{h,\hbar}^{\DHL,-\DHL}(z,\zbar) =\cr 
& ={{\Gamma({d\over 2}-1)}\over{\Gamma(\Delta_{L})\Gamma({\Delta_{L}-{d\over 2}+1})}}  (h\bar{h})^{\Delta_{L}+{d\over 2}}(h-\bar{h})(z\bar{z})^{{h+\bar{h}}\over 2}C_{h-\bar{h}}^{({d\over2}-1)}\Big( {{z+\bar{z}}\over{2\sqrt{z \bar{z}}}} \Big).
}}

\noindent Using the relations between coordinates $r,t$ and $z,\bar{z}$ it is easy to see that $(z\bar{z})^{{h+\bar{h}}\over 2}=e^{it(h+\bar{h})}$. Next, one can use the relation between Gegenbauer polynomials and hypergeometric functions:

\eqn\math{C_{n}^{(\alpha)}(z)={{(2\alpha)_{n}}\over{n!}}{}_2F_{1}(-n,2\alpha +n, \alpha + {1\over 2};{{1-z}\over{2}}),
}

\noindent which for $h-\bar{h}=l\gg 1$ gives:

\eqn\mathagain{C_{l}^{({d\over 2}-1)}\Big( {{z+\bar{z}}\over{2\sqrt{z \bar{z}}}} \Big) = {{l^{d-3}}\over{\Gamma{(d-2)}}} {}_2F_{1}(-l, l+d-2, {{d-1}\over{2}};{1\over{2}}-{1\over{2}}({{z+\bar{z}}\over{2\sqrt{z \bar{z}}}})).
}

\noindent With the help of the following properties of hypergeometric functions:

\eqn\proponetwo{\eqalign{
{}_2F_{1}(a,b,c; z)&=(1-z)^{-b}{}_2F_{1}(c-a, b, c; {z\over{z-1}}),\cr
\lim_{m,n\to \infty}&{}_2F_{1}(m,n,b;{z\over{m n}})={}_0F_{1}(b;z).
}}

Using these, together with the assumption that in the Regge limit the values of $x^{+}l$ and $x^{-}l$ are fixed constants: $x^{+}l=a_1$ and $x^{-}l=a_2$ while $l\to \infty$, one can easily see\foot{By noting that:
\eqn\Foot{\Gamma(x-{1\over{2}})=2^{2-2x}\sqrt{\pi}{{\Gamma(2x-1)}\over{\Gamma(x)}}.
}} 
that \final\ reproduces \iprr. This confirms the validity of the impact parameter representation.

\appendix{D}{Anomalous dimensions of heavy-light double-trace operators in $d=2$}
The OPE data of the heavy-light double trace operators  in $d=2$ dimensions can be directly obtained from the heavy-light Virasoro vacuum block \refs{\FitzpatrickVUA, \FitzpatrickZHA}. For completeness, in this appendix we investigate the anomalous dimension of $[\OH\OL]_{\hbar,h-\hbar}$ in $d=2$ following the discussion in Section 3. As in $d=4$, we introduce an impact parameter representation following \KulaxiziDXO. We calculate the anomalous dimensions to $\OO(\mu)$ by solving the crossing equation and then use the impact parameter representation to relate them to the bulk phase shift. We find a precise agreement between the two. Using the bulk phase shift we furthermore determine the anomalous dimension to second order in $\mu$. Much of the discussion follows closely the four-dimensional case and will be briefer. 

\subsec{Anomalous dimensions in the Regge limit using bootstrap}
The conformal blocks in two dimension are given by \refs{\DolanUT,\PolandEPD}
\eqn\BlocksTwoD{
  g_{\Delta,J}^{\Delta_{12},\Delta_{34}}(z,\zbar) = k_{\Delta+J}(z)k_{\Delta-J}(\zbar)+(z\lra\zbar)\,,
}
where $k_\beta(z)$ was defined in \KFunction. Similar to the four dimensional case, the blocks for heavy-light double-trace operators simplify in the heavy limit ($\DH\sim\CT$) 

\eqn\BlockTwoD{
  g_{[\OH\OL]_{h,\hbar}}^{\DHL,-\DHL}(z,\zbar)= (z\zbar)^{{1\over 2}(\DH+\DL)}(z^h\zbar^\hbar+(z\lra\zbar))\,.
}
Inserting this form of the conformal blocks in \OPESAnalyticallyContinued\ together with the OPE coefficients in the Regge limit \ReggeOpe\ and approximating the sums with integrals, one can due to symmetry extend the region of integration and it is easily found that the disconnected correlator in the T-channel is reproduced. 

Similar to the four-dimensional case the stress tensor dominates at order $\mu$ in the T-channel. The block of the stress tensor after analytic continuation in the Regge limit is given by 
\eqn\StressTensorBlockFourD{
  g_{T_{\mu\nu}} = {24i\pi e^{-\rho}\over \sigma}+\ldots\, ,
} 
where $\ldots$ denote non-singular terms. As in the four-dimensional case, this has to be reproduced in the S-channel by the term in \OPESAnalyticallyContinued\ proportional to $-i\pi\gamma$. 

With the conformal blocks \BlockTwoD, the imaginary part in the S-channel to  $\OO(\mu)$ is given by 
\eqn\ReggeBootTwoD{
  {\rm Im}(G(z,\zbar))|_{\mu}=-i\pi C_{\DL}\int_0^\infty dh \int_0^h d\hbar (h\hbar)^{\DL-1}\gamma^{(1)}(h,\hbar)\left(z^{h}\zbar^\hbar+z^\hbar\zbar^{h}\right).
}
Using the ansatz $\gamma^{(1)}(h,\hbar)=c_1h^a \hbar^b$ we find that the T-channel contribution is reproduced for $a=0$ and $b=1$ (see Appendix A.2 for details). We thus find using \OpeStressTensor\
\eqn\AnDimFirstOrderTwoD{\eqalign{
  \gamma^{(1)} = -{6\lambda_{\OH\OH T_{\mu\nu}}\lambda_{\OL\OL T_{\mu\nu}}\over \mu\DL}\hbar= -\hbar.
}}

To  $\OO(\mu^2)$ we can use \SecondOrderFourD\ to find the following contribution to the purely imaginary terms in the S-channel
\eqn\TwoDSecondOrder{
{\rm Im}(G(z,\zbar))|_{\mu^2}=-i\pi C_\DL\int_0^\infty dh\int_0^hd\hbar (h\hbar)^{\DL-1}\left(\gamma^{(2)}-{c_1^2\hbar\over 2}\right)(z^h\zbar^\hbar+z^\hbar\zbar^h).
}

\subsec{$2d$ impact parameter representation and relation to bulk phase shift}
Similar to the four-dimensional case we introduce an impact parameter representation in order to relate the anomalous dimension with the bulk phase shift. The impact parameter representation in $d=2$ is given by 
\eqn\IDefinitionTwoD{
  \II_{h,\hbar} \equiv C(\DL)\int_{M^+} d^2p (-p^2)^{\Delta-1} e^{-ipx} (h-\hbar)\delta(p\cdot \ebar+h+\hbar)\,\delta\left({p^2\over 4}+h \hbar\right)\, ,
}
with straightforward generalization of the $d=4$ case explained above. This is again chosen such that when the impact parameter represetation is integrated over $h,\hbar$ the disconnected correlator is reproduced: 
\eqn\IdefFreeCorr{
  \int_0^\infty dh\int_0^h \II_{h,\hbar} = {1\over [(1-z)(1-\zbar)]^\DL}.
}

The discussion of the phase shift is completely analogous to the four-dimensional case, as in \Relation\ we find the following relation between the bulk phase shift and the anomalous dimension to second order in $\mu$
\eqn\RelationTwoD{\eqalign{
  \gamma^{(1)}&=-{\delta^{(1)}\over \pi}\cr
  \tilde{\gamma}^{(2)}-{c_1^2p^-\over 4}&=-{\delta^{(2)}\over \pi}.
}}
In \KulaxiziDXO\ the phase shift in $d=2$ was found to be 
\eqn\PhaseShiftTwoD{\eqalign{
  \delta^{(1)} &= {\pi\over 2}\sqrt{-p^2}e^{-L}\cr 
  \delta^{(2)} &= {3\pi\over 8}\sqrt{-p^2}e^{-L}.
}}
Using the identification $p^+=2h$ and $p^-=2\hbar$ together with \PandRho\ we find for the anomalous dimension in the Regge limit
\eqn\AnomSecondOrderTwoD{\eqalign{
  \gamma^{(1)} &= -\hbar\cr 
  \gamma^{(2)} &= -{1\over 4}\hbar .
}}
We thus see that the first order result agrees with that obtained from bootstrap \AnDimFirstOrderTwoD. Furthermore, the second order correction agrees also in $d=2$ with the result (6.40) in \KulaxiziDXO.

\appendix{E}{Discussion of the boundary term integrals}

There are a few integrals containing total derivative terms that we have ignored throughout this paper and we analyze more carefully here. Let us start with a total derivative term which shows up in the real part of the correlator at $\OO(\mu)$. It is given by\foot{We are again using variables $n$ and $l$, one can notice that $n=\bar{h}$ and $l=h-\bar{h}$. It is trivial to prove that $\partial_{n}=\partial_{h}+\partial_{\bar{h}}$.}:

\eqn\boundaryfirst{I_1={1\over 2}(z\bar{z})^{-{1\over 2}(\Delta_{H}+\Delta_{L})}\int_{0}^{+\infty}dl \Big[ P^{(0)}\gamma^{(1)} g_{n+l,n}^{\Delta_{HL},-\Delta_{HL}}(z,\bar{z}) \Big]_{n=0}^{n\rightarrow  \infty}.
}

\noindent Let us focus on the integrand: $\Big[ P^{(0)}\gamma^{(1)} g_{n+l,n}^{\Delta_{HL},-\Delta_{HL}}(z,\bar{z}) \Big]_{n=0}^{n\rightarrow  \infty}$. When $n=0$, the expression within the brackets trivially vanishes. On the other hand, when $n\rightarrow \infty$, it takes the form $n^{2\Delta_{L}-2}(z\bar{z})^{n}\times f(l)$, where $f$ is some function of $l$ only. We are instructed here to take the limit $n\rightarrow \infty$ independently of all other limits (recall that the Regge limit is taken after the integration). For generic values $0<(z,\bar{z})<1$ it is clear that
$\lim_{n\rightarrow\infty}\Big[ P^{(0)}\gamma^{(1)} g_{n+l,n}^{\Delta_{HL},-\Delta_{HL}}(z,\bar{z}) \Big] =\lim_{n\rightarrow \infty} n^{2\Delta_{L}-2}(z\bar{z})^{n}\times f(l)\rightarrow 0$. In other words, the expression $\Big[ P^{(0)}\gamma^{(1)} g_{n+l,n}^{\Delta_{HL},-\Delta_{HL}}(z,\bar{z}) \Big]_{n=0}^{n\rightarrow  \infty} \rightarrow 0$, and we conclude that the integral \boundaryfirst\ does not contribute to the S-channel expansion of the correlator.

There are a few more integrals of similar kind that appear at $\OO(\mu^2)$. We will analyse one of them here:

\eqn\bndscnd{I_2={{-i\pi}\over 2}(z\bar{z})^{-{1\over 2}(\Delta_{H}+\Delta_{L})}\int_{0}^{+\infty}dl \Big[ P^{(0)}(\gamma^{(1)})^{2} g_{n+l,n}^{\Delta_{HL},-\Delta_{HL}}(z,\bar{z}) \Big]_{n=0}^{n\rightarrow  \infty}\, .
}

\noindent The same logic can be applied here. Again, the value of the expression in brackets at $n=0$ is trivially zero, while for large $n$ it behaves like: $n^{2\Delta_{L}+d-4}(z\bar{z})^{n} \tilde{f}(l)$. As long as $(z,\zbar)<1$, this vanishes exponentially in the limit $n\rightarrow \infty$. One concludes therefore that the integral \bndscnd\ vanishes. The same logic is valid for all other integrals of similar total derivative terms that appear at $\OO(\mu^2)$.

\appendix{F}{An identity for the bulk phase shift.}

\noindent The aim is to elaborate on the results of \KulaxiziDXO\  for the bulk phase shift in a black hole background as computed in gravity. Firstly, let us note the following identity involving hypergeometric functions:
\eqn\idhyperA{\eqalign{
\sum_{n=0}^\infty a(n) x^n& \,_2F_1[\tau_0+2 n+1,{d\over 2}-1,\tau_0+2 n-{d\over 2}+3,x]=\,_2F_1[\tau_0+1,{\tau_0\over 2},{\tau_0\over 2}+2,x]\cr
a(n)&={2^{2n}\over n!} {\tau_0+2\over\tau_0+2+2n}{({\tau_0\over 2}  +1-{d\over 2})_n  \left({\tau_0+1\over 2}\right)_n\over (\tau_0+n+2-{d\over 2})_n},\quad \tau_0\neq 0\,.
}}
Given that both sides of the equality can be expressed as an infinite series expansion around $x=0$, one simply needs to show that the expansion coefficients match to all orders in $x$. This is proven in Appendix G.

Consider now the case $\tau_0=k(d-2)$ where $k\in N^\star$. Setting $x\equiv e^{-2 L}$ and multiplying both sides with $e^{-[k(d-2)+1]L}$ yields:
\eqn\almostpirel{\eqalign{
&\Pi_{k(d-2)+1,k(d-2)+1} (L)=\sum_{n=0}^\infty \beta_n \Pi_{k(d-2)+2n+1,d-1}(L)\cr
&\beta(n)\equiv   \pi^{(1-k)(d-2)\over 2} \,{a(n)\over\left(k(d-2)+1\right)_n} \,{\Gamma\left[k(d-2)-{d\over 2}+2 n+3\right]\over\Gamma\left[{k(d-2)\over 2}+2\right]}\,. 
}}
The left hand side represents the hyperbolic space propagator for a scalar field of squared mass equal to $k(d-2)+1$ in a hyperbolic space of dimensionality $k(d-2)+1$ and is proportional to the $k$-th order expression for the bulk phase shift computed in gravity in \KulaxiziDXO, where
\eqn\bpsbh{\delta^{(k)}(S,L) ={1\over k!}{2\Gamma\left({dk+1\over 2}\right)\over \Gamma\left({k(d-2)+1\over 2}\right)} {\pi^{1+{k(d-2)\over 2}}\over\Gamma\left({k(d-2)\over 2}+1\right)  }S\,\Pi_{k(d-2)+1,k(d-2)+1}(L)\,.} 

On the other hand, the right-hand side of \almostpirel\ expresses the $k$-th order term of the bulk phase shift as an infinite sum of $(d-1)$-dimensional hyperbolic space propagators for fields with mass-squared equal to $m^2=k(d-2)+1+2n$.

It can be shown \refs{\SimmonsDuffinUY, \KulaxiziIXA} that the analytically continued T-channel scalar conformal block in the Regge limit behaves like:
 \eqn\ReggeBlock{
 g_{\Delta,J}(\sigma,\rho)=i\,c_{\Delta,J}{\Pi_{\Delta-1,d-1}(\rho)\over \sigma^{J-1}}\,,
 }
 where
\eqn\cdef{
c_{\Delta, J}={4^{\Delta+J-1} \Gamma\left({\Delta+J-1\over 2}\right)\Gamma\left({\Delta+J+1\over 2}\right)\over \Gamma({\Delta+J\over 2})^2} \, {2\Gamma\left(\Delta-{d\over 2}+1\right)\over \pi^{1-{d\over 2}} \Gamma\left({\Delta-1}\right)} \,.
} 
Here $\Pi_{\Delta-1,d-1}$ denotes as usual the $(d-1)$-dimensional hyperbolic space propagator for a massive scalar of mass-squared $m^2=(\Delta-1)$.

It follows that the $k$-th order term in the $\mu$-expansion of the bulk phase shift in a black hole background can be expressed as an infinite sum of conformal blocks corresponding to operators of twist $\tau=\tau_0(k)+2 n=k(d-2)+2 n$ and spin $J=2$ in the Regge limit. In other words, we can write:
\eqn\deltaksum{\eqalign{
i\,\delta^{(k)}(S,L)&=  \,\, f(k)\,\,\sum_{n=0}^\infty \lambda_k(n) \,\, g_{\scriptscriptstyle \tau_0(k)+2n+2,2}^{R}(S,L)  \cr
\lambda_k(n)&=a(n) \, {2^{-4 n}\left[\left({\tau_0(k)+4\over 2}\right)_n\right]^2\over \left({\tau_0(k)+3\over 2}\right)_n\left({\tau_0(k)+5\over 2}\right)_n} ,\quad \tau_0(k)=k(d-2)\,
}}
where
\eqn\defak{f(k)\equiv {\sqrt{\pi}\over 64} \,\, {1\over 2^{k(d-2)} \,\,k!}  \, {\Gamma\left({k d+1\over 2}\right)\Gamma\left({k(d-2)+4\over 2}\right)\over\Gamma\left({k(d-2)+5\over 2}\right)\Gamma\left({k(d-2)+3\over 2}\right)}\,,
}
and
\eqn\ReggeFourier{g_{\scriptscriptstyle \Delta,J}^{R}(S,L)=i c_{\Delta,J}\,S^{J-1}\,\Pi_{\Delta-1,d-1}(L)\,.
}
%where we substituted $\mu$ in terms of the CFT data according to \OpeStressTensor.

\appendix{G}{An identity for hypergeometric functions.}

\noindent Here we will show that for $q\neq 0$,
\eqn\idhyperA{\eqalign{
\sum_{n=0}^\infty a(n) x^n \,_2F_1[q+2 n+1,{d\over 2}-1,q+2 n-{d\over 2}+3,x]&=\,_2F_1[q+1,{q\over 2},{q\over 2}+2,x]\cr
a(n)={2^{2n}\over n!} {q+2\over q+2+2n}{({q\over 2}  +1-{d\over 2})_n  \left({q+1\over 2}\right)_n\over (q+n+2-{d\over 2})_n},\quad q\neq 0\,.
}}
Given that both sides of the equality can be expressed as an infinite series expansion around $x=0$, one simply needs to show that the expansion coefficients match to all orders in $x$. 
Let us first set:
\eqn\bcdef{\eqalign{
b(n,m)&\equiv {1\over m!}{\left(q+1+2n\right)_m \left({d\over 2}-1\right)_m\over \left(q-{d\over 2}+2 n+3\right)_m}\,\cr
c(\ell)&\equiv {1\over\ell !} {\left(q+1\right)_\ell \left({q\over 2}\right)_\ell\over \left({q\over 2}+2\right)_\ell} ={(q+1)_\ell\over\ell !} {q(q+2)\over (q+2\ell)(q+2\ell+2)}\,,
}}
such that:
\eqn\hypers{\eqalign{ \,_2F_1[q+2 n+1,{d\over 2}-1,q+2 n-{d\over 2}+3,x]&=\sum_{m=0}^\infty b(n,m)x^m,\cr
\,_2F_1[q+1,{q\over 2},{q\over 2}+2,x]&=\sum_{\ell=0}^\infty c(\ell)x^\ell.
}}
It is easy to check that the coefficients of the first few powers of $x$ precisely match. Indeed, e.g.,
\eqn\fcheck{\eqalign{
a(0) b(0,0)-c(0)&=0\cr
a(1)b(1,0)+a(0)b(0,1)-c(1)&=0 \cr
a(2)b(2,0)+a(1)b(1,1)+a(0)b(0,2)-c(2)&=0.
}}
To show that the above identity is true for all powers of $x$ we must show that:
\eqn\ida{\sum_{k=0}^\ell a(k)b(k,\ell-k)=c(\ell)\,,
}
for all $\ell\in N$. 
The left-hand side of \ida\ can be easily summed to yield:
\eqn\idb{
\sum_{k=0}^\ell a(k)b(k,\ell-k)={1\over \ell !}{\Gamma[q+1+\ell]\over\Gamma[q]} {(q+2)\over (q+2\ell)(2+2\ell+q)}\,,
}
which can be trivially shown to be equal to $c(\ell)$.

\listrefs

\bye